\documentclass[11pt]{article}
\pdfoutput=1

\usepackage[dvipsnames]{xcolor}
\usepackage{graphicx}
\usepackage{subcaption}
\usepackage{epsfig}
\usepackage{epstopdf}
\usepackage[T1]{fontenc}

\DeclareSymbolFont{myletters}{OML}{ztmcm}{m}{it}
\DeclareMathSymbol{\uplambda}{\mathord}{myletters}{"15}

\usepackage{amsfonts}

\usepackage{hyperref}
\hypersetup{
    colorlinks=true,
    linkcolor=blue!70,
    citecolor=RedViolet,
    filecolor=magenta,      
    urlcolor=cyan,
}

\usepackage{color}
\usepackage{float}
\usepackage{multirow}
\usepackage[toc,page]{appendix}

\usepackage{lmodern} 
\usepackage{amsmath}                                                    
\usepackage{amsthm}                                                     
\usepackage{amssymb}
\usepackage{amsfonts}
\usepackage{mathptmx}
\usepackage{slashed}
\usepackage{latexsym}

\numberwithin{equation}{section} 

\newcommand{\newc}{\newcommand}
\newc{\be}{\begin{equation}}
\newc{\ee}{\end{equation}}
\newc{\bg}{\begin{gathered}}
\newc{\eg}{\end{gathered}}
\newc{\tref}[1]{Table \ref{#1}}
\newc{\eref}[1]{Equation \eqref{#1}}
\newc{\su}[1]{$SU(#1)$}
\newc{\bm}[1]{\mathbf{#1}}
\newc{\fref}[1]{Figure \ref{#1}}

\newc{\ra}{\rightarrow}
\newc{\lra}{\leftrightarrow}
\newc{\ov}{\overline}
\newc{\ba}{\begin{eqnarray}}
\newc{\ea}{\end{eqnarray}}
\newc{\mf}{\mathsf}
\newc{\nn}{\nonumber}

\begin{document}

\begin{titlepage}
\thispagestyle{empty}

                \vspace*{0.7cm}

                \begin{center}
                                                { {\bf \Large{ On the LHC signatures of $SU(5)\times U(1)'$ F-theory motivated models}}}
                        \\[12mm]
                        A. Karozas$^{a}$~\footnote{E-mail: \texttt{akarozas@uoi.gr}}, 
                G. K. Leontaris$^{a}$~\footnote{E-mail: \texttt{leonta@uoi.gr}},
                        I. Tavellaris$^{a}$~\footnote{E-mail: \texttt{i.tavellaris@uoi.gr}}, N. D. Vlachos$^{b}$~\footnote{E-mail: \texttt{vlachos@physics.auth.gr}}
                \end{center}
                \vspace*{0.50cm}
                        \centerline{$^{a}$ \it
                                Physics Department, University of Ioannina}
                        \centerline{\it 45110, Ioannina,        Greece}
                \vspace*{0.2cm}
                \centerline{$^{b}$ \it
                        Department of Nuclear and Elementary Particle Physics, Aristotle University of Thessaloniki}
                \centerline{\it 54124, Thessaloniki,    Greece}
                \vspace*{1.20cm}
                \begin{abstract}

                        We study  low energy implications of F-theory GUT models based on  $SU(5)$   extended by a $U(1)'$ symmetry which  couples non-universally to the three families of quarks and leptons.  This gauge group arises naturally from the maximal exceptional gauge symmetry of an elliptically fibred internal space, at a single  point of  enhancement, $E_8\supset SU(5)\times SU(5)'\supset SU(5)\times U(1)^4.$
                        Rank-one fermion mass textures and a tree-level top quark coupling are guaranteed  by imposing  a $Z_2$ monodromy group  which identifies two abelian factors of       the above breaking sequence.    The $U(1)'$ factor of the gauge symmetry is an anomaly free linear  combination of the three remaining abelian symmetries left over by $Z_2$. Several classes of models       are obtained, distinguished with respect to the $U(1)'$         charges of the representations, and possible extra zero modes coming in vector-like representations.  The predictions of these models are investigated and are compared with the LHC results and  other related experiments. Particular cases interpreting the B-meson anomalies observed in LHCb and   BaBar experiments are also discussed.

                \end{abstract}
        \end{titlepage}

\thispagestyle{empty}
\vfill
\newpage
{
  \hypersetup{linkcolor=black}
\tableofcontents
}

\section{Introduction}
Despite its tremendous success, the Standard Model (SM) of the strong and electroweak  interactions leaves many theoretical questions unanswered.
Accumulating evidence  of the last few decades indicates that  new ingredients are required in order to describe  various  New Physics (NP) phenomena 
 in particle physics and cosmology.   Amongst other shortcomings, the minimal SM spectrum does  not accommodate a dark matter candidate particle
  and  the tiny neutrino masses cannot be naturally incorporated. Regarding this latter issue, in particular,  an elegant   
way to interpret the tiny  masses of the three neutrinos and their associated  oscillations, is the seesaw mechanism~\cite{Mohapatra:1979ia}  which
brings into the scene  right-handed neutrinos and a new (high) scale. 
Interestingly, this scenario fits nicely inside the framework of (supersymmetric)  grand unified theories (GUTs) 
which unify the three fundamental forces at a high (GUT) scale. 
Besides, several ongoing neutrino experiments suggest the existence of a 
  `sterile' neutrino  which could also be a suitable dark matter candidate \cite{Boser:2019rta, Boyarsky:2018tvu}. 
   Many other lingering questions  regarding the existence of possible remnants of a covering theory, such as leptoquarks, 
  vectorlike families, supersymmetry signatures and neutral gauge bosons,  are 
 expected to find an answer in the experiments carried out at the Large Hadron Collider (LHC).
Remarkably,  many field theory GUTs incorporate most of the above novel  fields into larger representations, 
while, after spontaneous symmetry breaking of the initial gauge symmetry takes place,  cases where additional $U(1)$ factors  
 survive down to low energies implying masses for  the associated neutral gauge  bosons accessible to ongoing experiments. 
However, while GUTs with the aforementioned   new features are quite appealing, they come at a price. Various extra fields, including heavy gauge bosons and other colored states, contribute to fast proton decay and other rare processes.

In contrast to plain field theory GUTs,  string theory  alternatives are subject to selection rules 
and other restrictions, while new  mechanisms are operative  which, under certain conditions, could eliminate 
 many of the problematic states and  undesired  features.      F-theory models \cite{Vafa:1996xn, Beasley:2008dc, Beasley:2008kw},
  in particular, appear to  naturally include such attractive  features which  are attributed to the 
 intrinsic geometry of the compactification manifold and the fluxes piercing 
 matter curves where the various  supermultiplets reside. In other words, the
geometric properties and the fluxes can be chosen so that, among other things, 
  determine   the desired symmetry breaking, reproduce the known multiplicity of the
 chiral fermion  families, and eliminate the colored triplets in  Higgs representations. 
Moreover, in F-theory constructions, the gauge symmetry of the resulting effective field theory model 
is  determined in terms of the geometric structure of the elliptically fibred  internal compactification space. 
In particular, the non-abelian  part of the gauge symmetry is associated with the codimension-one singular  fibers,
while possible abelian and discrete symmetries are determined in terms of the Mordell-Weil (MW) and Tate-Shafarevish (TS) groups.~\footnote{For a recent survey  see for example~\cite{Weigand:2018rez}. For earlier F-theory reviews see~\cite{Heckman:2010bq,Leontaris:2012mh,Maharana:2012tu}.
For models with Mordell-Weil $U(1)$'s and other issues see \cite{Morrison:2012ei}-\cite{Kimura:2019qxf}.} 
 For elliptically fibred manifolds, the non-abelian gauge 
symmetry is a simply laced  algebra (i.e. of type $A,D$ or $E$ in Lie classification), the highest 
one corresponding to the exceptional group of $E_8$. At fibral singularities, certain divisors wrapped with 7-branes are associated with
subgroups of $E_8$, and are interpreted as the GUT group of the effective theory. In addition, $U(1)$ symmetries 
may accompany the non-abelian group.  The origin of the latter could emerge either  from the $E_8$-part commutant to the GUT
 group or  from MW and TS groups mentioned above. Among the  various possibilities, there is a particularly 
 interesting case where a neutral gauge boson $Z'$ associated with some abelian factor with non-universal couplings
 to the quarks and leptons, obtains mass at the TeV region. Since the SM gauge bosons  
couple universally  to quarks (and leptons) of the three families, the existence of non-universal couplings would lead to 
deviations from SM predictions that could be interpreted as an indication for NP beyond the SM.

Within the above context, in~\cite{Romao:2017qnu} a first systematic study of 
a generic class of F-theory semi-local models based on the $E_8$ subgroup  $SU(5)\times U(1)'$ 
has been presented~\footnote{For similar works on anomaly free $U(1)'$s see also~\cite{Ellis:2017nrp,Allanach:2018vjg}.}.
 The anomaly-free $U(1)'$   symmetry  has non-universal couplings to the three
chiral  families and the  corresponding gauge boson receives  a low energy (a few TeV) mass. In that work, some 
particular properties of representative examples  were examined 
 in connection with new flavour phenomena and  in particular, the B-meson physics explored 
 in LHCb~\cite{Aaij:2014ora, Aaij:2017vbb, Aaij:2019wad}.
In the present  work we extend the previous analysis by performing a systematic investigation  into  the 
various predictions and the constraints imposed on
all possible classes of viable models emerging from this framework. Firstly we distinguish classes of models with respect 
to  their low energy spectrum and  properties under the $U(1)'$ symmetry. 
We find a class of models with a minimal MSSM spectrum at low energies. The members of this 
class are differentiated by the charges under the additional $U(1)'$.
 A second class of anomaly free viable 
effective low energy models, contains additional MSSM multiplets
coming in vector-like pairs.  In the present work, we analyse the constraints imposed by various processes 
on the list of models of the first class.
The phenomenological analysis of a characteristic example  containing extra vector-like states is also presented, while 
the complete analysis of these models is postponed  for a future publication.
In the first category  (i.e. the minimal models),  anomaly cancellation conditions 
impose non-universal $Z'$ couplings to the three fermion field families.  As a result, in most cases,
 the  stringent bounds coming from  kaon decays imply a relatively large $Z'$ gauge boson mass that lies beyond
 the accessibility of the present day experiments.  
On the contrary, models with extra vetor-like pairs offer a variety of possibilities. There 
are viable cases where the fermions of the first two generations are characterised by the same $Z'$ couplings.
In such cases, the stringent bounds of the $K-\bar K$ system can be evaded and a $Z'$ mass can be as
low as a few TeV.

  The  work is organised and presented in five sections. In section \ref{sec2} we start by
 developing the general formalism of a $Z'$ boson coupled non-universally to MSSM.
 Then, we discuss flavour violating processes in the quark and lepton sectors,
 putting emphasis on  contributions to  B-meson anomalies and other
 deviations from the SM  explored in LHC and other related experiments. (To make
 the paper self contained, all relevant recent experimental bounds
 are also given). In section \ref{sec3} we start with a brief review of 
 local F-theory GUTs. Then, using generic properties of the compactification manifold and the flux
 mechanism, we apply  well defined rules and spectral cover 
 techniques to construct viable effective models.
 We  concentrate on a $SU(5)\times U(1)'$ model embedded in $E_8$ and impose anomaly
 cancellation conditions to  obtain a variety of  consistent F-theory effective models. We distinguish between two categories; a class of models  with a  MSSM (charged) spectrum (possibly with some extra neutral singlet fields) and a second one where the MSSM spectrum is extended with vector-like  quark and charged lepton representations. 
  In section \ref{sec4} we analyse the phenomenological implications of the 
  first class, paying particular attention to B-meson physics 
  and lepton flavour violating decays. Some consequences of 
  the models with extra vector-like fields are discussed in section \ref{sec5}, while a detailed investigation 
  into the whole class of models will be presented in a future publication. 
  In section \ref{sec6} we present  our conclusions.  Computational 
  details are given in the appendix.

        \section{Non-universal $Z^{\prime}$ interactions}\label{sec2}

In the Standard Model, the neutral gauge boson  couplings to fermions 
with the same electric charge are equal, therefore, the corresponding tree-level interactions  are flavour diagonal.  
However, this is not always true in  models with additional $Z^{\prime}$ bosons  associated with extra 
$U(1)'$  factors emanating from higher symmetries. If the $U(1)'$ charges of all or some of the three fermion 
families are different, significant flavour mixing effects might occur even at tree-level. In this section we review some basic facts 
about non-universal $U(1)$'s and develop the necessary formalism to be used subsequently. 

\subsection{Generalities and formalism}

To set the stage, we first   consider the neutral part of the Lagrangian including the $Z^{\prime}$ interactions with fermions
in the gauge eigenstates basis \cite{Langacker:2008yv, Langacker:2000ju} :

\be  \label{nc_lagrange}
-\mathcal{L}_{NC}\supset{eJ_{EM}^{\mu}A_{\mu}+\frac{g}{c_{W}}J^{(0)\;\mu}Z^{0}_{\mu}+g^{\prime}J^{\prime\;\mu}Z^{\prime}_{\mu}}~,
\ee
\noindent where $A_{\mu}$ is the massless photon field, $Z^{0}$ is the  neutral gauge boson of the SM  and $Z^{\prime}$ is the new  boson associated with the extra $U(1)^{\prime}$ gauge symmetry. Also $g$ and $g^{\prime}$ are the gauge couplings of the weak $SU(2)$ gauge symmetry and the new $U(1)^{\prime}$ symmetry respectively. For shorthand, we have denoted $\cos{\theta_{W}}$ $(\sin{\theta_{W}})$ as $c_{W}$ ($s_{W}$)  where $\theta_{W}$ is the weak mixing angle
with $g=e/\tan{\theta_{W}}$. The neutral current associated with the $Z^{\prime}$  boson can be written as:
\be\label{nc1}
J^{\prime\;\mu}=\bar{f_{L}^{0}}\gamma^{\mu}q^{\prime}_{f_{L}}f_{L}^{0}+\bar{f_{R}^{0}}\gamma^{\mu}q^{\prime}_{f_{R}}f_{R}^{0}~,
\ee

\noindent where $f_{L}^{0}$ ($f_{R}^{0}$) is a column vector of left (right) chiral fermions of a given type ($u$, $d$, $e$ or $\nu$) in the gauge  basis and $q^{\prime}_{f_{L,R}}$ are diagonal $3\times{3}$ matrices of $U(1)^{\prime}$ charges. $f_{L}$ denotes  chiral fermions in the mass eigenstates basis,  related to gauge eigenstates via unitary transformations of the form
\be  \label{nc12} 
f_{L}^{0}=V_{f_{L}}^{\dagger}f_{L}\; ,\quad f_{R}^{0}=V_{f_{R}}^{\dagger}f_{R}~\cdot
\ee

\noindent $V_{f_{L,R}}$ are unitary matrices responsible for the diagonalization of the Yukawa matrices $Y_{f}$,
\be 
Y_{f}^{diag}=V_{f_{R}}Y_{f}V_{f_{L}}^{\dagger}~, 
\ee

\noindent with the CKM matrix defined by the combination: 
$$V_{CKM}=V_{u_{L}}V_{d_{L}}^{\dagger}~.$$

 In the mass eigenbasis, the neutral current \eqref{nc1} takes the form :

\be\label{nc2}
J^{\mu}= \bar{f_{L}}\gamma^{\mu}\mathit{Q}^{\prime}_{f_{L}}f_{L}+\bar{f_{R}}\gamma^{\mu}\mathit{Q}^{\prime}_{f_{R}}f_{R}
\ee 

\noindent where 
\be \label{mixingchargematrix}
\mathit{Q}^{\prime}_{f_{L}}\equiv V_{f_L}q^{\prime}_{f_{L}}V_{f_L}^{\dagger}\; , \;\;  \mathit{Q}^{\prime}_{f_{R}}\equiv V_{f_R}q^{\prime}_{f_{R}}V_{f_R}^{\dagger}\; .
\ee 

If the $U(1)^{\prime}$ charges in the $q^{\prime}_{f_{L}}$ matrix are equal, then $q^{\prime}_{f_{L}}$ is the unit matrix up to a common charge factor and due to the unitarity of $V_{f}$'s the current in \eqref{nc2} becomes flavour diagonal. For models  with family non-universal $U(1)^{\prime}$ charges, the mixing matrix  $\mathit{Q}^{\prime}_{f_{L}}$ is non-diagonal and flavour violating terms appear in the effective theory.

\subsection{Quark sector flavour violation}

\subsubsection{$b\rightarrow{sl^{+}l^{-}}$ and $R_{K}$ anomalies}

The possible existence of non-universal $Z^{\prime}$ couplings to fermion families, may lead to departures from 
the SM predictions and leave clear  signatures in present day or near future  experiments. Such contributions 
strongly depend on the mass $M_{Z^{\prime}}$ of the $Z^{\prime}$ gauge boson, the $U(1)^{\prime}$ gauge coupling, $g^{\prime}$, the  $U(1)^{\prime}$ fermion charges and  the mixing  matrices $V_{f}$. 
A particularly interesting case reported by LHCb~\cite{Aaij:2019wad} and BaBar\cite{Lees:2015ymt} collaborations, indicate that there may be anomalies observed in B-meson decays,
associated with the transition  $b\rightarrow{sl^{+}l^{-}}$, where $l=e,\mu,\tau$.
 Current LHCb measurements of $b$ decays to different lepton pairs hint to
 deviations from lepton universality.  In particular, the analysis performed in the $q^2$ invariant mass
 of the lepton pairs in the range 
 $1.1$ GeV${}^2< q^2<6$ GeV${}^2$ for the ratio of the branching ratios $Br(B\rightarrow{K^{(*)}\ell^{+}\ell^{-}}),\ell=\mu,e$ gives~\cite{Aaij:2019wad}
\begin{equation}
R_{K}\equiv\frac{Br(B\rightarrow{K^{}\mu^{+}\mu^{-}})}{Br(B\rightarrow{K^{}e^{+}e^{-}})}
\simeq{0.846}^{+0.016\,\rm (stat)}_{-0.014\,\rm (syst)}\label{BKme} ~,
\end{equation}
where statistical and systematic uncertainties are indicated. Similarly, the results for $B\to K^*(892)\ell^+\ell^-$
(where $K^*\to K\pi$), for the same ratio~(\ref{BKme}) are found to be 
$R_{K^{*}}\simeq{0.69}$. Since  the SM strictly  predicts $R_{K^{(*)}}^{SM}=1$, these  results strongly suggest that NP scenarios  where  lepton universality is violated should be explored. In  the case  $l=\mu$ in particular, additional experimental and theoretical arguments suggest that NP may be related with the muon channel  \cite{Aaij:2015oid,Khachatryan:2015isa,Altmannshofer:2013foa}.

\begin{figure}[t]
\centering
\includegraphics[width=1.0\textwidth]{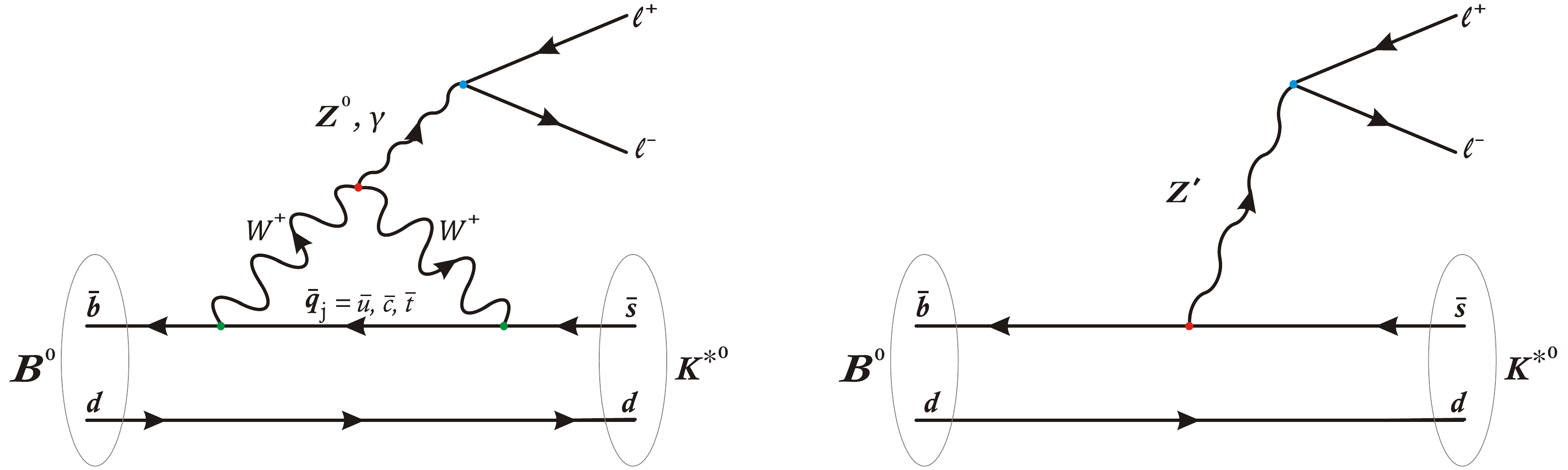}
\caption{\small{Left panel: Example of a Feynman diagram contributing to $B^{0}\rightarrow{K^{*}l^{+}l^{-}}$ in the SM context. Right panel: Tree level contribution in models with non-universal $Z^{\prime}$'s. }}\label{fig:feyn1}
\end{figure}

In the SM, $B\rightarrow{K^{(*)}l^{+}l^{-}}$ can  only be realised  at the one-loop level involving $W^{\pm}$ flavour changing interactions (see left panel of Figure \ref{fig:feyn1}). However, the existence of a $Z^{\prime}$ (neutral) gauge boson bearing non-universal couplings to fermions,  can lead to  tree-level contributions (right panel of Figure \ref{fig:feyn1}) which might explain (depending on the model) the observed anomalies.

The effective Hamiltonian describing the interaction is given by~\cite{Altmannshofer:2013foa}

\be  \label{Heff}
H_{eff}^{b\rightarrow{sll}}=-\frac{4G_{F}}{\sqrt{2}}\frac{e^2}{16\pi^{2}}(V_{tb}V_{ts}^{*})\sum_{k=9,10}\left(C_{k}^{ll}\mathcal{O}_{k}^{ll}+C_{k}^{\prime ll}\mathcal{O}_{k}^{\prime ll}\right)
\ee
where the symbols $ \mathcal{O}^{xx}_{n}$ stand for the following dimension-6 operators,
\begin{align*}
\mathcal{O}^{ll}_{9}&=(\bar{s}\gamma^{\mu}P_{L}b)(\bar{l}\gamma_{\mu}l),\; \;\;\; \mathcal{O}^{ll}_{10}=(\bar{s}\gamma^{\mu}P_{L}b)(\bar{l}\gamma_{\mu}l)            \\
\mathcal{O}^{\prime ll}_{9}&=(\bar{s}\gamma^{\mu}P_{R}b)(\bar{l}\gamma_{\mu}\gamma_{5}l),\; \; \mathcal{O}^{\prime ll}_{9}=(\bar{s}\gamma^{\mu}P_{R}b)(\bar{l}\gamma_{\mu}\gamma_{5}l) ~,          
\end{align*}
 and $C_{k}$  are Wilson coefficients displaying the strength of the interaction. Also, in \eqref{Heff}, $G_{F}$ is the Fermi coupling constant and $V_{tb}$, $V_{ts}^{*}$ are elements of the CKM matrix.

The latest data for $R_{K^{(*)}}$ ratios can be interpreted by assuming a negative contribution to the Wilson coefficient $C_{9}^{\mu\mu}$, while 
all the other Wilson coefficients\footnote{Alternative scenarios suggest : $C_{10}^{\mu\mu}\approx{0.73\pm{0.14}}$ or $C_{9}^{\prime\mu\mu}=-C_{10}^{\prime\mu\mu}\approx{-0.53\pm{0.09}}$.} should be negligible, or vanishing \cite{Aebischer:2019mlg,Alok:2019ufo,Alguero:2019ptt,Kowalska:2019ley,Arbey:2019duh}. 
The current best fit value is $C_{9}^{\mu\mu}\approx{-0.95\pm{0.15}}$.

 In the presence of a non-universal $Z^{\prime}$ gauge boson,
 the  $C_{9}^{\mu\mu}$ Wilson coefficient is given by :

\be \label{c9wilson}
C_{9}^{\mu\mu}=-\frac{\sqrt{2}}{4G_{F}}\frac{16\pi^{2}}{e^2}\left(\frac{g^{\prime}}{M_{Z^{\prime}}} \right)^{2}\frac{(Q^{\prime}_{d_{L}})_{23}(Q^{\prime}_{e_{L}})_{22}}{V_{tb}V_{ts}^{*}}\; .
\ee
 The desired value for the $C_{9}$ coefficient could be achieved by appropriately tuning the ratio $g^{\prime}/M_{Z^{\prime}}$. However, large suppressions may occur from the matrices $Q^{\prime}_{f}$. In any case, the predictions must not create conflict with well known bounds coming from rare processes such as the mixing effects in neutral meson systems.

\subsubsection{Meson mixing}

\begin{figure}[t]
\centering
\includegraphics[width=1.0\textwidth]{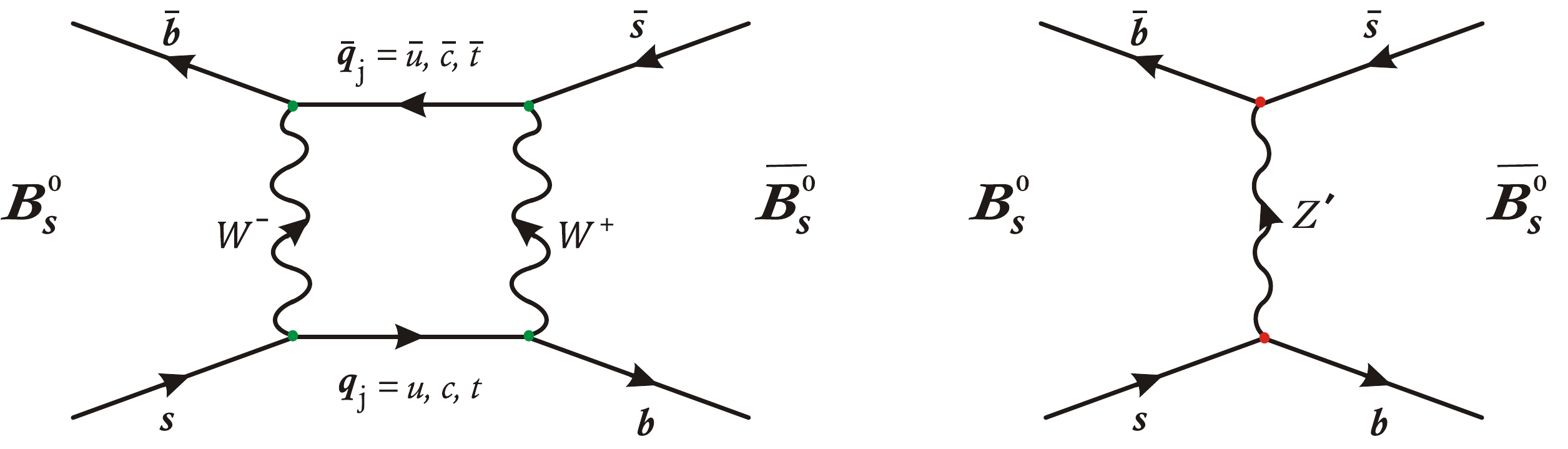}
\caption{\small{Left figure: Representative \emph{box} diagram contribute to $(B_{s}^{0}-\bar{B_{s}^{0}})$ mixing in the SM. Right figure: Tree level contribution in models with non-universal $Z^{\prime}$ gauge bosons. }}\label{fig:BBmixfeynman}
\end{figure}

Flavor changing $Z^{\prime}$ interactions in the quark sector can also induce significant contributions  to the mass splitting in a neutral meson system. A representative example is given in Figure \ref{fig:BBmixfeynman}. The diagrams show contributions to $B_{s}^{0}[s\bar{b}]$ mixing in the SM (left) and tree-level contributions in  non-universal $Z^{\prime}$ models (right). 

For a meson $P^{0}$ with quark structure $[q_{i}\bar{q}_{j}]$, the contribution from $Z^{\prime}$ interactions to the mass splitting is given by \cite{Langacker:2000ju}:

\be \label{eq:mesonmix}
\Delta M_{P}\simeq 4\sqrt{2}G_{F}M_{P}F_{P}^{2}\left(\frac{M_{W}}{g\cdot{c_w}}\right)^{2}\left(\frac{g^{\prime}}{M_{Z^{\prime}}}\right)^{2}\frac{1}{3}\mathrm{Re}[(Q^{\prime}_{q_{L}})_{ij}^{2}]
\ee
 
\noindent where $M_{W}$ is the mass of the $W^{\pm}$ gauge bosons and $M_{P}$, $F_{P}$ is the mass and the decay constant of the meson $P^{0}$ respectively.

There are large uncertainties in the SM computations of $\Delta M_{P}$, descending especially from QCD factors and the CKM matrix elements. Nevertheless, the experimental results suggest that there is still some room for NP contributions.

Next, we review  theoretical and experimental constraints for  $P^{0}-\bar{P^{0}}$ meson systems to   be taken into account  in what follows.

$\bullet$ $B_{s}^{0}-\bar{B^{0}_{s}}$ mixing:

\noindent  $B_s$ mixing can be described by the effective Lagrangian 

\be  
\mathcal{L}^{NP}=-\frac{4G_{F}}{\sqrt{2}}(V_{tb}V_{ts}^{*})^{2}[C_{bs}^{LL}(\bar{s}_{L}\gamma_{\mu}b_{L})^{2}+h.c.]~,
\ee

\noindent where $C_{bs}^{LL}$ is a Wilson coefficient which modifies the SM prediction as follows \cite{DiLuzio:2017fdq}:

\be \label{eq:bbmixbound1}
\Delta M_{s}^{pred}=|1+C_{bs}^{LL}/R_{SM}^{loop}|\Delta M_{s}^{SM}~,
\ee

\noindent with $R_{SM}^{loop}=1.3397\times{10^{-3}}$. 

A model with non-universal $Z^{\prime}$ couplings to fermions induces the following Wilson coefficient:

\be \label{eq:bbmixwilson} 
C_{bs}^{LL}=\frac{\eta^{LL}}{4\sqrt{2}G_{F}}\left( \frac{g^\prime}{M_{Z^{\prime}}}\right)^{2}\frac{(Q^{\prime}_{d_{L}})^{2}_{23}}{(V_{tb}V_{ts}^{*})^{2}} 
\ee

\noindent where $\eta^{LL}\equiv{\eta^{LL}(M_{Z^{\prime}})}$ is a constant which encodes renormalisation group effects. This constant has a weak dependence\footnote{For $M_{Z^{\prime}}\in{[1,10]}$ TeV it turns out that $\eta^{LL}\in{[0.79,0.75]}$, see \cite{Buras:2000if, DiLuzio:2017fdq}.} on the $M_{Z^{\prime}}$ scale. In our analysis we  consider that $\eta^{LL}=0.79$ which corresponds to $M_{Z^{\prime}}=1\; TeV$.

For the SM contribution $\Delta M_{s}^{SM}$ we consider the result obtained in Ref. \cite{King:2019lal},

\begin{equation*}
\Delta M_{s}^{SM}=(18.5^{+1.2}_{-1.5})\; ps^{-1}~,
\end{equation*}
\noindent which when compared with the experimental bound \cite{Tanabashi:2018oca}, $\Delta M_{s}^{exp}=(17.757^{+0.021}_{-0.021})$ $ps^{-1}$,  shows through eq. \eqref{eq:bbmixbound1}, that a small positive $C_{bs}^{LL}$ is allowed.

$\bullet$ $K^{0}-\bar{K^{0}}$ mixing :
\\
SM computations for the mass split in the neutral Kaon system are a combination of short-distance and long-distance effects, given as \cite{Buras:2014maa}

\be \label{SMkkmix}
\Delta M_{K}^{SM}=(0.8\pm{0.1})\Delta M_{K}^{Exp}~,
\ee

\noindent where the experimental data are given by \cite{Tanabashi:2018oca}:

\[\Delta M_{K}^{exp}\simeq{3.482}\times{10^{-15}}\;\; GeV. \]
This small discrepancy between SM computations and experiment can be explained by including  NP effects into the analysis. Thus, according to \eqref{SMkkmix},  the contribution of a non-universal $Z^{\prime}$ boson  to $\Delta M_{K}$ must satisfy the following constraint \cite{Chen:2018dfc};

\begin{equation} \label{eq:kaonconstraint}
\Delta M_{K}^{NP}\lesssim{0.2\times{\Delta M_{K}^{exp}}}~,
\end{equation}
 \noindent where $\Delta M_{K}^{NP}$ can be computed directly from the formula \eqref{eq:mesonmix}.

$\bullet$ $D^{0}-\bar{D^{0}}$ mixing:
\\
Neutral $D$ mesons  consist of up-type quarks, $D^{0}:\rightarrow{[c\bar{u}]}$. The experimental measurements for $D^{0}-\bar{D^{0}}$ oscillations are sensitive to the ratio:

\be \label{eq:ddmixratio}
x_{D}\equiv{\frac{\Delta M_{D}}{\Gamma_{D}}}~,
\ee

\noindent where $\Gamma_{D}$ is the total decay width  of $D^{0}$  and the observed value for the ratio is $x_{D}\simeq{0.32}$ \cite{Amhis:2016xyh}. Since the process is subject to large theoretical and experimental uncertainties, we will simply consider NP contributions to $x_{D}$  less or equal to the experimental value.

\subsubsection{Leptonic Meson Decays : $P^{0}\rightarrow{l_{i}\bar{l}_{i}}$}

 In the SM the decay of a neutral meson $P^{0}$ into a lepton ($l_{i}$) and its anti-lepton ($\bar{l_{i}}$) is realised at the one-loop level. While in the SM these processes are suppressed due to  GIM~\cite{Glashow:1970gm} cancellation mechanism, in non-universal $Z^{\prime}$ models substantially larger  tree-level contributions  may be allowed. The decay width induced by  $Z^{\prime}$ interactions can be written in terms of the SM decay $P^{-}\rightarrow{l_{i}\bar{\nu_{i}}}$ as \cite{Langacker:2000ju}:
 
 \be \label{eq:Ptolili}
 \Gamma(P^{0}\rightarrow{l_{i}\bar{l_{i}}})\simeq{ 8\frac{\Gamma(P^{-}\rightarrow{l_{i}\bar{\nu_{i}}})}{|V^{CKM}_{kj}|^{2}}\frac{M_{P}^{3}\sqrt{M_{P}^{2}-4m_{l_{i}}^{2}}}{(M_{P}^{2}-m_{l_{i}}^{2})^{2}} \left(\frac{g^{\prime}}{M_{Z^{\prime}}}\right)^{4}\left(\frac{M_{Z_{0}}}{g}\right)^{4} |(Q^{\prime}_{q_{L}})_{mn}(Q^{\prime}_{e_{L}})_{ii}|^{2} }~,
 \ee 

\noindent where the indices $j,k$ refer to the quark structure $[q_{j}\bar{q_{k}}]$ of the meson $P^{-}$ appearing in the SM interaction. Similarly, the indices $m,n$ are used here to denote the quark structure of the neutral meson $P^0$. All the relevant experimental bounds for this type of interactions can be found in \cite{Tanabashi:2018oca}.

\subsection{Lepton flavour violation}

\subsubsection{$P^{0}\rightarrow{l_{i}\bar{l_{j}}}$}

The lepton flavour violation process $P^{0}\rightarrow{l_{i}\bar{l_{j}}}$ is similar to the previous one  where $i=j$.  The decay width due to tree-level $Z^{\prime}$ contributions  is given by \cite{Langacker:2000ju}:

 \be 
 \Gamma(P^{0}\rightarrow{l_{i}\bar{l_{j}}})\simeq{ 4\frac{\Gamma(P^{-}\rightarrow{l_{i}\bar{\nu_{i}}})}{|V^{CKM}_{kr}|^{2}} \left(\frac{g^{\prime}}{M_{Z^{\prime}}}\right)^{4}\left(\frac{M_{Z_{0}}}{g}\right)^{4} |(Q^{\prime}_{u_{L}})_{mn}(Q^{\prime}_{e_{L}})_{ij}|^{2}} \cdot
 \ee 

\noindent As previously, the indices $k, r$ are used to denote the quark structure $[q_{r}\bar{q_{k}}]$ of the meson participating in the SM interaction, while generation indices $m,n$ refer to the quark structure of $P^{0}$. Bounds and predictions for these rare interactions will be given in the subsequent analysis.

\subsubsection{$(g-2)_{\mu}$}

The anomalous magnetic moment of the muon $a_{\mu}\equiv{(g-2)/2}$, is  measured with high accuracy. However there exists a discrepancy between experimental measurements and precise SM computations \cite{Tanabashi:2018oca}:
\be 
\Delta a_{\mu}\equiv{a_{\mu}^{exp}- a_{\mu}^{SM}}=261(63)(48)\times{10^{-11}}
\ee

\noindent where $a_{\mu}^{SM}=116 591 830(1)(40)(26)\times{10^{-11}}$.

\begin{figure}[t]
\centering
\includegraphics[width=1.\textwidth]{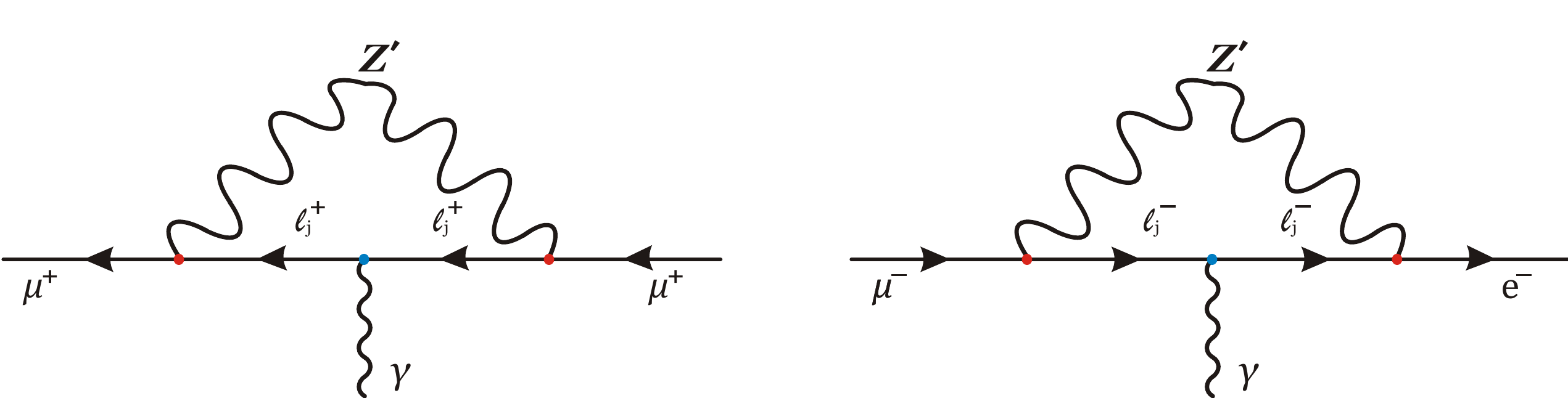}
\caption{\small{Left side: Contribution of a non-universal $Z^{\prime}$ boson into the magnetic moment of (anti)muon. Right side: Contribution to the decay, $\mu^{-}\rightarrow{e^{-}\gamma}$. Any of the three (anti)leptons ($j=e,\mu ,\tau$) could run in to the loop due to the non-universal charges under the extra $U(1)$ symmetry.}}\label{gminus2}
\end{figure}

This difference  may  be explained by NP contributions. In the case of a $Z^{\prime}$ neutral boson, loop diagrams like the one shown on the left side of Figure \ref{gminus2} contribute to $\Delta a_{\mu}$. Collectively, the 1-loop contribution from a non-universal $Z^{\prime}$ bosons is \cite{Jegerlehner:2009ry}:

\be\label{g2contribution} 
\Delta a_{\mu}^{Z^{\prime}}=-\frac{m_{\mu}^{2}}{8\pi^{2}}\left(\frac{g^{\prime}}{M_{Z^{\prime}}}\right)^{2}\sum_{j=1}^{3}|(Q^{\prime}_{e_{L}})_{2j}|^{2}F(x_{l_{j}}^{Z^{\prime}})
\ee

\noindent where $x_{l_{j}}^{Z^{\prime}}:=(m_{l_{j}}/M_{Z^{\prime}})^{2}$ with the loop function defined as:

\be
F(x)=\frac{5x^{4}-14x^{3}+39x^{2}-38x-18x^{2}\ln(x)+8}{12(1-x)^{4}} \cdot
\ee

\noindent In our analysis we will consider that $\Delta a_{\mu}^{Z^{\prime}}$ must be less or equal to  $\Delta a_{\mu}$.

\subsubsection{$l_{i}\rightarrow{l_{j}\gamma}$}

A flavour violating $Z^{\prime}$ boson contributes also to radiative decays of the form $l_{i}\rightarrow{l_{j}\gamma}$. The 1-loop diagram of the strongly constrained decay $\mu^{-}\rightarrow{e^{-}\gamma}$ is displayed in Figure \ref{gminus2} (right). Considering only $Z^{\prime}$ contributions, the branching ratio for this type of interactions is given by \cite{Lavoura:2003xp}:
\be\label{eq:ltolgammaBR}
\mathrm{Br}(l_{i}\rightarrow{l_{j}\gamma})=\frac{e^{2}}{16\pi\Gamma_{l_{i}}}\left( m_{l_{i}}-\frac{m_{l_{j}}^{2}}{m_{l_{i}}}\right)^{3}(g^{\prime})^{2}\sum_{f}\left[y_{2}(Q^{\prime}_{e_{L}})_{fj}(Q^{\prime}_{e_{L}})_{fi}\right]~,
\ee
\noindent where the index $f=1,2,3$ refers to the lepton running inside the loop, $\Gamma_{l_{i}}$ is the total decay width of the lepton $l_{i}$ and $y_{2}$ is a loop function that can be found in \cite{Lavoura:2003xp}. The most recent experimental bounds are:
\begin{equation*}  
\mathrm{Br(\mu\rightarrow{e\gamma})}< 4.2\times{10^{-13}},\; \mathrm{Br(\tau\rightarrow{e\gamma})}< 3.3\times{10^{-8}}\;\;\mathrm{and}\;\;\;\mathrm{Br(\tau\rightarrow{\mu\gamma})}< 4.4\times{10^{-8}} ~\cdot 
\end{equation*}
\noindent Dominant constraints are expected to come from the muon decay.

\subsubsection{$l_{i}\rightarrow{l_{j}l_{k}\bar{l}_{j}}$}

A lepton flavour violating $Z^{\prime}$ boson mediates (at tree-level) three-body leptonic decays of the form $l_{i}\rightarrow{l_{j}l_{j}\bar{l}_{k}}$. The branching ratio is given by \cite{Okada:1999zk}:

\be \label{eq:eldecayBR1}
\mathrm{Br}(l_{i}\rightarrow{l_{j}l_{j}\bar{l}_{k}})=\frac{m_{l_{i}}^{5}}{768\pi^{3}\Gamma_{l_{i}}}\left(\frac{g^{\prime}}{M_{Z^{\prime}}}\right)^{4}|(Q^{\prime}_{e_{L}})_{ij}(Q^{\prime}_{e_{L}})_{kj}|^{2}~,
\ee

\noindent where the masses of the produced leptons have been neglected.

For decays of the form $l_{i}\rightarrow{l_{j}l_{k}\bar{l}_{j}}$ with $k\neq{j}$ the branching ratio is

\be \label{eq:eldecayBR2}
\mathrm{Br}(l_{i}\rightarrow{l_{j}l_{j}\bar{l}_{k}})=\frac{m_{l_{i}}^{5}}{1536\pi^{3}\Gamma_{l_{i}}}\left(\frac{g^{\prime}}{M_{Z^{\prime}}}\right)^{4}|(Q^{\prime}_{e_{L}})_{ik}(Q^{\prime}_{e_{L}})_{jj}+(Q^{\prime}_{e_{L}})_{ij}(Q^{\prime}_{e_{L}})_{jk}|^{2}~.
\ee

\noindent The dominant constraint comes from the muon decay $\mu^{-}\rightarrow{e^{-}e^{-}e^{+}}$, with branching ratio bounded as $Br(\mu\rightarrow{eee})<10^{-12}$ at $90\%$ confidence level \cite{Bellgardt:1987du}.

\section{Non-universal $U(1)^{\prime}$ models from F-theory}\label{sec3}

We  now turn on to the class of F-theory constructions accommodating abelian factors bearing non-universal couplings with 
 the three families of the Standard Model. 
 As  already mentioned, we focus on constructions based on an elliptically fibred compact space with $E_8$ being the maximal singularity, and assume a divisor in the internal manifold where the associated non-abelian  gauge symmetry  is $SU(5)$. 
With  this choice,  $E_8$ decomposes as 
 \be 
 E_8\supset SU(5)\times SU(5)_{\perp}~\cdot\label{SU5square}
 \ee 
  We will restrict our analysis in local constructions and describe the resulting effective theory
 in terms of the Higgs bundle picture which makes use of the adjoint scalars where only the Cartan generators acquire a non-vanishing 
 vacuum expectation value (VEV)\footnote{For non-diagonal generalisations (\emph{T-branes}) see \cite{Cecotti:2010bp}.}.
 In the local  picture we may work with the spectral data (eigenvalues  and eigenvectors) which, for the case of $SU(5)$,
 are associated with the $5^{th}$ degree polynomial
  \be\label{sc1}
 \mathcal{C}_{5}=\sum_{k=0}^5 b_kt^{5-k}=b_{0}t^{5}+b_{1}t^{4}+b_{2}t^{3}+b_{3}t^{2}+b_{4}t+b_{5} =0~.
 \ee
This defines the spectral cover for the fundamental representation of 
$SU(5)$. Furthermore, as is the case for any $SU(n)$,  the five roots
\be   Q= \{t_1, t_2, t_3, t_4, t_5\},\label{Tweights}
\ee
must add up to zero,
\be 
-b_1\equiv \sum_{i=1}^5 t_i=0~.\label{5roots}
 \ee 
 The remaining coefficients are generically non-zero, $b_k\ne 0, k=0,2,3,4,5$ and carry the geometric properties 
 of the internal manifold. 
 
The zero-mode spectrum of the effective low energy theory  descends from the decomposition of the $E_8$ adjoint.
With respect to the  breaking pattern (\ref{SU5square}), it decomposes as follows:
\ba
248&\ra& (24,1)+\boxed{(1,24)+(10,5)+(\overline{5},10)+(5,\overline{10})}+(\overline{10},\overline{5})\,.\label{E825}
\ea
Ordinary  matter and Higgs fields, including the appearance of possible singlets  in the spectrum, appear in the box of the right-hand side in~(\ref{E825}) and  transform in bi-fundamental representations, with respect to the two $SU(5)$s. 
From the above, we observe that the GUT  decuplets transform in the fundamental of  $SU(5)_{\perp}$, whilst the
$\bar 5, 5$-plets  are in the antisymmetric representation of the `perpendicular' symmetry. For our present 
purposes however, it is adequate to work in the limit where the perpendicular symmetry 
reduces down to the Cartan subalgebra according to the 
breaking pattern $SU(5)_{\perp}\to U(1)^4_{\perp}$. In this picture,
the GUT representations are characterised by the appropriate combinations of the five  weights given in~(\ref{Tweights}).    The five 10-plets 
in particular, are counted by $t_{1,2,\dots 5}$ and the fiveplets
which originally transform as decuplets under the second $SU(5)_{\perp}$ 
are characterised by the ten combinations $t_i+t_j$. 
 In the geometric description, 
it is said that the $SU(5)$ GUT representations reside in Riemann surfaces (dubbed matter curves $\Sigma_a$) formed by the intersections   of the $SU(5)$ GUT divisor with `perpendicular' 
7-branes. These properties are summarised in the following notation
\be  
\Sigma_{10_{t_i}}:\; 10{_{t_i}}, \ov{10}{_{-t_i}},\;\;  \Sigma_{5_{t_i+t_j}}:\; \ov{5}_{t_i+t_j}\,,  5_{-t_i-t_j},\;\;
\Sigma_{1_{t_i-t_j}}:\; 1_{t_i-t_j} ~\cdot  \label{Treps}
\ee 
As we have seen above, since the weights $t_{i=1,2,3,4,5}$ associated with the $SU(5)_{\perp}$ group, are the roots of the  polynomial~(\ref{sc1}), they can be expressed as functions of 
the coefficients  $b_{k}$'s which carry the information
regarding the  geometric properties of the compactification
manifold. Based on this fact,
in the subsequent analysis, we will make use of the topological 
invariant quantities and flux data to determine the spectrum and
the parameter space of  the effective low energy models under 
consideration.

We start by determining the zero-mode spectrum of the possible
classes of models within the context discussed above. According 
to the spectral cover description, see equations~(\ref{sc1}-\ref{Treps}), 
the various matter curves of the theory accommodating the $SU(5)$ GUT multiplets are determined  by the following equations: 

\be
\Sigma_{10_{t_i}}: \; \; P_{10}:= b_{5}\sim{\prod_{i=1}^{5}{t_{i}}}=0 ~,
\ee

\noindent and 

\be
 \Sigma_{5_{t_i+t_j}}: \; \; P_{5}:= b_{3}^{2}b_{4}- b_{2}b_{3}b_{5}+b_{0}b_{5}^{2}\sim{\prod_{i\neq{j}}{(t_{i}+t_{j})}}=0~. 
\ee
If all five roots $t_i$ of the polynomial (\ref{sc1}) are distinct 
and expressed as holomorphic functions of the coefficients $b_k$,
then, simple counting shows that there can be five matter curves 
accommodating the  tenplets(decuplets) and ten matter curves 
where the fiveplets(quintuplets) can reside. This would imply
that the  polynomial (\ref{sc1}) could be expressed as 
a product $\prod_{i=1}^{5}(\alpha_i t_{i}+\beta_i)$, with the
coefficients $\alpha_i, \beta_i$  carrying the topological properties of the manifolds, while 
being in the same field as the
original $b_k$. However, in the generic case not all five solutions $t_i(b_k)$ belong to the same field with $b_k$. In effect, there are  
monodromy   relations among subsets of 
the roots $t_{i}$, reducing  the number of independent matter
curves. Depending on the specific geometric properties  of the
compactification manifold,  we can have a variety
of  factorisations of the spectral cover polynomial $\mathcal{C}_{5}$.
(The latter are parametrised  by the Cartan subalgebra modulo the Weyl group $W(SU(5)_{\perp})$).
In other words, generic solutions  imply branch 
cuts and some  roots are indistinguishable.  The simplest case is when two of them are subject
to a $Z_2$ monodromy, 
\be  
Z_2:\, t_1=t_2~.\label{Z2M}
\ee 
Remarkably, there is an immediate  implication of the $Z_2$ monodromy  in the  effective field theory model.
It allows the  tree-level coupling in the superpotential
\be 
10_{t_1}10_{t_2} 5_{-t_1-t_2} \stackrel{Z_2}\longrightarrow  10_{t_1}10_{t_1} 5_{-2t_1}~,
\ee 
which can induce a heavy  top-quark mass as required by low energy phenomenology.

Returning to the spectral cover description, under 
the $Z_2$ monodromy,  the polynomial (\ref{sc1}) 
is factorised accordingly to
\be
\label{sc2}
\mathcal{C}_{5}=({a_1+a_2t+a_3t^2})(a_4+a_7t)(a_5+a_8t)(a_6+a_9t)~,
\ee
 where the existence of the second degree polynomial is not
 factorisable in the sense presented above,
 indicating thus, 
 that the corresponding roots $t_1,t_2$ are connected by $Z_2$. 
 
Comparing this with the spectral polynomial in (\ref{sc1}), we can  extract the relations between the coefficients $b_{k}$ and $a_{j}$. 
 Thus, one gets
 
\ba\label{bfroma}
b_0&=&a_3 a_7 a_8 a_9
\; ,\nn\\
b_1&=&a_3 a_6 a_7 a_8+a_3 a_4 a_9 a_8+a_2 a_7 a_9 a_8+a_3 a_5 a_7 a_9\; ,\nn\\
b_2&=&a_3 a_5 a_6 a_7+a_2 a_6 a_8 a_7+a_2 a_5 a_9 a_7+a_1 a_8 a_9 a_7+a_3 a_4 a_6 a_8+a_3 a_4 a_5 a_9+a_2 a_4
   a_8 a_9\; ,\nn\\
   b_3&=&a_3 a_4 a_5 a_6+a_2 a_5 a_7 a_6+a_2 a_4 a_8 a_6+a_1 a_7 a_8 a_6+a_2 a_4 a_5 a_9+a_1 a_5 a_7 a_9+a_1 a_4
   a_8 a_9\; ,\nn\\
   b_4&=&a_2 a_4 a_5 a_6+a_1 a_5 a_7 a_6+a_1 a_4 a_8 a_6+a_1 a_4 a_5 a_9\; ,\nn\\
   b_5&=&a_1 a_4 a_5 a_6~\; .
   \ea
We impose the $SU(5)$ constraint $b_1=0$  assuming the \emph{Ansatz} \cite{Dudas:2010zb} 
\[ a_2 =-c ( a_6 a_7 a_8+a_5 a_7 a_9+ a_4 a_8 a_9),\; a_3=c  a_7 a_8 a_9~,\]

\noindent where a new holomorphic section $c$ has been introduced. Substituting into (\ref{bfroma}) one gets
\ba \label{bfromac}
b_0&=&c\; a_7^2 a_8^2 a_9^2 \; ,\nn \\
b_2&=&a_9 \left(a_1 a_7 a_8-\left(a_5^2
   a_7^2+a_4 a_5 a_8 a_7+a_4^2 a_8^2\right) a_9 c\right)-c a_6^2 a_7^2 a_8^2 -c a_6 a_7 \left(a_5 a_7+a_4 a_8\right) a_9 a_8 
\; ,\nn\\
b_3&=&a_1 \left(a_6 a_7 a_8+\left(a_5 a_7+a_4 a_8\right) a_9\right)-\left(a_5 a_7+a_4 a_8\right) \left(a_6
   a_7+a_4 a_9\right) \left(a_6 a_8+a_5 a_9\right) c\; ,\\
b_4&=&a_1 \left(a_4 a_6 a_8+a_5 \left(a_6 a_7+a_4 a_9\right)\right)-a_4 a_5 a_6 \left(a_6 a_7 a_8+\left(a_5
   a_7+a_4 a_8\right) a_9\right) c\; ,\nn\\
b_5&=&a_1 a_4 a_5 a_6~\; .\nn
\ea

The equations of tenplets and fiveplets can now be expressed in terms of the holomorphic sections 
$a_{j}$'s and $c$. In the case of the tenplets we end up with four factors
\be \label{tens}
P_{10}=a_{1}\times a_{4}\times a_{5}\times a_{6}~,
\ee
\noindent which correspond to four matter curves accommodating the tenplets of $SU(5)$. 
Substitution of (\ref{bfromac}) in to $P_5$ factorises the equation into seven factors  corresponding  to seven distinct fiveplets
\ba \label{fives}
\nn
P_{5}&=&\left(a_5 a_7+a_4 a_8\right)\times \left(a_6 a_7+a_4 a_9\right)\times \left(a_6 a_8+a_5 a_9\right)\\
 & &\times\left(a_6 a_7
   a_8+a_4 a_9 a_8+a_5 a_7 a_9\right)\times \left(a_1-a_5 a_6 a_7 c-a_4 a_6 a_8 c\right)\\ \nn
  & &\times\left(a_1-a_5 a_6 a_7 c-a_4
   a_5 a_9 c\right)\times \left(a_1-a_4 a_6 a_8 c-a_4 a_5 a_9 c\right) ~.  
   \ea

Finally, we  compute the homologies of the section $a_{j}$'s and $c$, and consequently of each matter curve. This can be done by using the known homologies of the  $b_{k}$ coefficients: 
\be
[b_{k}]=(6-k)\mathsf{c}_1-t =\eta-k\,\mathsf{c}_1
\ee
\noindent where  $\mathsf{c}_1$ is the $1^{st}$ Chern class of the tangent bundle to $S_{GUT}$, $-t$ the $1^{st}$  Chern class of the normal bundle to  $S_{GUT}$ and $\eta =6\,\mathsf{c}_1-t$. The homologies of $a_{j}$'s and $c$ are presented in Table \ref{a_homologies}, while the homologies of the various matter curves are given in Table \ref{mc_homologies}. Because there are more $a$'s than $b$'s, three homologies  which are taken to be  $[a_7]=\chi_{7}$, $[a_8]=\chi_{8}$ and $[a_9]=\chi_{9}$, remain unspecified.

\begin{table}
\begin{center}
 \resizebox{\textwidth}{!}{ 
\begin{tabular}{cccccc|ccc|c}
\hline\hline
$a_1\quad$& $a_2\quad$& $a_3\quad$& $a_4\quad$& $a_5\quad$& $a_6\quad$& $a_7\quad$&$a_8\quad$&$a_9\quad$& c
\\
$\eta-2\mathsf{c}_1-\chi\quad$& $\eta-\mathsf{c}_1-\chi\quad$& $\eta-\chi\quad$& $-\mathsf{c}_1+\chi_7\quad$& $-\mathsf{c}_{1}+\chi_8\quad$& $-\mathsf{c}_1+\chi_9\quad$& $\chi_7\quad$&$\chi_8\quad$&$\chi_9\quad$&$\quad\eta-2\chi$\\
\hline\hline
\end{tabular}}
\end{center}
\caption{\small{Homology classes of the coefficients $a_j$ and $c$. Note that $\chi=\chi_{5}+\chi_{7}+\chi_9$ where $\chi_{7},\chi_{8},\chi_9$ are the unspecified homologies of the coefficients $a_5$, $a_7$ and $a_9$ respectively.}}
\label{a_homologies}
\end{table}

\begin{table}
\begin{center}
 \resizebox{\textwidth}{!}{ 
\begin{tabular}{l||cccc|ccccccc}
\hline\hline
Matter Curve & $\Sigma_{10_1}$ & $\Sigma_{10_2}$ &  $\Sigma_{10_3}$ &  $\Sigma_{10_4}$ &  $\Sigma_{5_1}$ &  $\Sigma_{5_2}$ & $\Sigma_{5_3}$ & $\Sigma_{5_4}$ & $\Sigma_{5_5}$ & $\Sigma_{5_6}$ & $\Sigma_{5_7}$ 
\\
Weights & $\pm{t_1}$ & $\pm{t_2}$ & $\pm{t_3}$ & $\pm{t_4}$ & $\pm{2t_1}$ & $\pm{(t_{1}+t_{3})}$ & $\pm{(t_{1}+t_{4})}$ & $\pm{(t_{1}+t_{5})}$ & $\pm{(t_{3}+t_{4})}$ & $\pm{(t_{3}+t_{5})}$ & $\pm{(t_{4}+t_{5})}$\\
Def. equation & $a_1$ & $a_4$ & $a_5$ & $a_6$ & $a_{6}a_{7}a_{8}+...$ & $a_{1}-...$ & $a_{1}-...$ & $a_{1}-...$ & $a_{5}a_{7}+...$ & $a_{6}a_{7}+...$ & $a_{6}a_{8}+...$ \\
Homology & $\eta-2\mathsf{c}_{1}-\chi$ & $\chi_{7}-\mathsf{c}_{1}$ & $\chi_{8}-\mathsf{c}_{1}$ & $\chi_{9}-\mathsf{c}_{1}$ & $\chi-\mathsf{c}_{1}$ & $\eta-2\mathsf{c}_{1}-\chi$ & $\eta-2\mathsf{c}_{1}-\chi$ & $\eta-2\mathsf{c}_{1}-\chi$ & $\chi_{7}+\chi_{8}-\mathsf{c}_{1}$ & $\chi_{7}+\chi_{9}-\mathsf{c}_{1}$ & $\chi_{8}+\chi_{9}-\mathsf{c}_{1}$ \\\hline\hline
\end{tabular}}
\end{center}
\caption{\small{Matter curves along with their $U(1)_{\perp}$ weights  ($\pm$ refer to $10/\overline{10}$ and $\bar 5/5$ respectively), their defining equation and the corresponding homology class.}}
\label{mc_homologies}
\end{table}

\subsection{$SU(5)\times{U(1)^{\prime}}$ in the spectral cover description }       

Our aim is to examine $SU(5)\times{U(1)^{\prime}}$  models and particularly the r\^ole of 
the non-universal  $U(1)'$ which should be consistently embedded in the covering group $E_8$.
Clearly, the $U(1)'$ symmetry should be  a linear combination
of the abelian factors residing in $SU(5)_{\perp}$. 
A convenient abelian basis to express the desired $U(1)'$ emerges in the following sequence of symmetry breaking
\ba
E_8
& \supset & E_6\times SU(3)_{\perp} \supset E_6\times U(1)_\perp \times{U(1)_{\perp}'} \label{E60} \\
& \supset & SO(10)\times U(1)_{\psi}\times U(1)_\perp \times{U(1)_{\perp}'} \label{SO(10)} \\
& \supset & SU(5)_{GUT} \times  U(1)_{\chi}  \times U(1)_{\psi} \times  U(1)_\perp \times{U(1)_{\perp}'} . \label{SU(5)0}
\ea

Then, the  Cartan generators corresponding to the four  $U(1)$'s are expressed as:
\begin{equation}
\label{4gen}
\begin{split}
Q_{\perp}'&=\frac{1}{2}
{\rm diag}(1,-1,0,0,0), 
\\
Q_{\perp}&=\frac{1}{2\sqrt{3}}{\rm diag}(1,1,-2,0,0),\\
Q_{\psi}&=\frac{1}{2\sqrt{6}}{\rm diag}(1,1,1,-3,0),\\
Q_{\chi}&=\frac{1}{2\sqrt{10}}{\rm diag}(1,1,1,1,-4).
\end{split}
\end{equation}

 The  monodromy $t_1\leftrightarrow t_2$  imposed in the previous section,  eliminates the 
 abelian factor corresponding to 
 $Q_{\perp}'$ with $t_1\ne t_2$.
Then we are left with  the remaining three $SU(5)_\perp$ generators
\ba
Q_{\perp},\;
Q_{\psi},\;
Q_{\chi}~,
\ea
given in (\ref{4gen}).
Next, we assume that a  low energy $U(1)^\prime$ is generated by a linear combination of the unbroken $U(1)$'s:
\be
Q^{\prime }= c_1 Q_\perp + c_2 Q_\psi + c_3 Q_\chi~\cdot\label{Qprime}
\ee
Regarding the coefficients $c_{1}, c_{2}, c_{3} $  the following normalisation condition       will be assumed
\be\label{norm_condition}
c_1^2+c_2^2+c_3^2 =1~,
\ee
while, further constraints will be  imposed by applying anomaly cancellation conditions.

\subsection{The Flux mechanism }
We now turn into the symmetry breaking procedure. In F-theory, fluxes are 
  used to generate the observed chirality  of the massless spectrum. Most precisely, we may consider two distinct classes of fluxes. Initially, a flux is introduced along a $U(1)_{\perp}$ and
 its geometric restriction along a specific matter  curve $\Sigma_{n_j}$ is parametrised with
 an integer number. Then, the  chiralities  of the $SU(5)$ representations are given by 
 \ba
  \# 10_i-\# \overline{10}_i&=& m_{i}\label{M5ii}\\
 \# 5_j-\# \overline{5}_j&=& M_{j}\label{M10jj}
 \ea

\noindent The integers $M_{i},m_j$ are subject to the chirality condition

\be\label{chiralitycondition}
\sum_{i}m_{i}=-\sum_{j}M_{j}=3
\ee

\noindent which coincides with the $SM$ anomaly conditions \cite{Marsano:2010sq,Palti:2012dd}

\noindent
Next, a flux in the direction of hypecharge, denoted as $\mathcal{F}_Y$, is turned on in order to break the $SU(5)_{GUT}$ down to the SM gauge group. This "hyperflux" is also responsible for the splitting of $SU(5)$ representations. If some integers   $N_{i,j}$ represent hyperfluxes piercing  
certain  matter curves, then the combined effect of the two type of fluxes into the 10-plets and 5-plets  is described according to:
\ba
{10}_{t_{j}}=
\left\{\begin{array}{ll}
        n_{{(3,2)}_{\frac 16}}-n_{{(\bar 3,2)}_{-\frac 16}}&=\;m_{j}\\
        n_{{(\bar 3,1)}_{-\frac 23}}-n_{{(
                        3,1)}_{\frac 23}}&=\;m_{j}-N_j\\
        n_{(1,1)_{+1}}-n_{(1,1)_{-1}}& =\;m_{j}+N_j\\
\end{array}\right.\,,
\label{F10j}
\ea
\ba
{5}_{t_{i}}=
\left\{\begin{array}{ll}
        n_{(3,1)_{-\frac 13}}-n_{(\bar{3},1)_{+\frac 13}}&=\;M_{i}\\
        n_{(1,2)_{+\frac 12}}-n_{(1,2)_{-\frac 12}}& =\;M_{i}+N_i\,\; .\\
\end{array}\right.
\label{F5i}
\ea

\noindent We note in passing that since the Higgs field is accommodated on a  matter curve of  type~(\ref{F5i}),
 an elegant solution to the doublet-triplet splitting problem is realised. Indeed,  imposing $M_i=0$ 
 the colour triplet is eliminated, while choosing  $ N_{i}\neq{0}$ we ensure the existence of massless doublets in the low energy spectrum.  
 
 \noindent 
 The $U(1)_Y$ flux is subject to the conditions 

\be\nn
\mathcal{F}_{Y}\cdot{\eta}=\mathcal{F}_{Y}\cdot{\mathsf{c}_{1}}=0~,
\ee

\noindent in order  to  avoid a heavy Green-Schwarz mass for the corresponding gauge boson. Furthermore, assuming  $\mathcal{F}_{Y}\cdot{\chi_{i}}=N_{i}$ (with $i=7,8,9$) and correspondingly $\mathcal{F}_{Y}\cdot{\chi}=N$, with $N=N_{7}+N_{8}+N_{9}$, we can find the effect of hyperflux on each matter curve. While $m_{i}$ and $M_{j}$ are subject to the constraint \eqref{chiralitycondition}, hyperflux integers $N_{7,8,9}$ are related to the undetermined 
homologies $\chi_{7,8,9}$ and as such, they are free parameters of the theory. The flux data and the SM content of each matter curve are presented in Table \ref{Spectable}. The particle content of  the matter curves arises from the decomposition of $10+\overline{10}$ and $5+\overline{5}$ pairs 
which reside on the appropriate matter curves.  
The MSSM chiral fields arise from the decomposition of $10$ and $5$,  and 
are denoted by $Q,L,u^c,d^c, e^c$.  Depending on the choice of the flux parameters, it is also possible that some of  
their conjugate fields appear in the light spectrum (provided of course that there are only three chiral families in the effective theory).
These conjugate fields arise from $\overline{10}$ and $\overline{5}$ and in Table \ref{Spectable} and  are denoted by 
 $\overline{Q}, \overline{L}, \overline{u^c},\overline{d^c},\overline{e^c}$.

In the same table we have also included the charges of the remaining $U(1)'$ symmetry.  We observe that the charges are functions of the $c_{1,2,3}$  coefficients which can be computed by applying anomaly cancellation conditions.

\begin{table}[t!]
 \resizebox{\textwidth}{!}{     
                \begin{tabular}{c|c|c|c|c}
                        \hline\hline
                        Matter Curve  &           $Q^\prime$        &    $N_Y$    &       M     &        SM Content      
                        \\ \hline
                                $\Sigma_{10_{1,\pm t_1}}$    &  $\frac{10 \sqrt{3} c_1+5 \sqrt{6} c_2+3 \sqrt{10} c_3}{60}$   & $-N$ & $m_{1}$  &   $m_{1}Q+(m_{1}+N) u^c +(m_{1}-N)e^c$   
                        \\
                        $\Sigma_{10_{2,\pm t_3}}$   &   $\frac{-20 \sqrt{3} c_1+5 \sqrt{6} c_2+3 \sqrt{10} c_3}{60}$    &    $N_7$    & $m_{2}$  &       $m_{2}Q+(m_{2}- N_7) u^c +(m_{2}+ N_7)e^c$        
                        \\
                        $\Sigma_{10_{3,\pm t_4}}$    &            $\frac{\sqrt{10} c_3-5 \sqrt{6} c_2}{20} $  &    $N_8$    & $m_{3}$  &   $m_{3}Q+(m_{3}- N_8) u^c +(m_{3}+N_8)e^c$       
                        \\
                        $\Sigma_{10_{4,\pm t_5}}$     &     $-\sqrt{\frac{2}{5}} c_3$   &    $N_9$    & $m_{4}$  &   $m_{4}Q+(m_{4}- N_9) u^c +(m_{4}+N_9)e^c$       
                        \\ \hline
                        $\Sigma_{5_{1,(\pm 2t_1)}}$   &     $-\frac{c_1}{\sqrt{3}}-\frac{c_2}{\sqrt{6}}-\frac{c_3}{\sqrt{10}}$     & $N$  & $M_{1}$ & $M_{1}\overline{d^c} + (M_{1}+N)\overline{L}$ 
                        \\
                        $\Sigma_{5_{2,\pm (t_1+t_3)}} $ &    $\frac{5 \sqrt{3} c_1-5 \sqrt{6} c_2-3 \sqrt{10} c_3}{30} $    & $-N$ &  $M_{2}$  &     $M_{2}\overline{d^c} + (M_{2}-N)\overline{L}$    
                        \\
                        $\Sigma_{5_{3,\pm (t_1+t_4)}}$  &    $-\frac{c_1}{2 \sqrt{3}}+\frac{c_2}{\sqrt{6}}-\frac{c_3}{\sqrt{10}}$     & $-N$ &  $M_{3}$  &     $M_{3}\overline{d^c} + (M_{3}- N)\overline{L}$    
                        \\
                        $\Sigma_{5_{4,\pm (t_1+t_5)}}$ & $\frac{-10 \sqrt{3} c_1-5 \sqrt{6} c_2+9 \sqrt{10} c_3}{60} $ & $- N$ &  $M_{4}$  &     $M_{4}\overline{d^c} + (M_{4}-N)\overline{L}$    
                        \\
                        $\Sigma_{5_{5,\pm (t_3+t_4)}} $ &     $\frac{c_1}{\sqrt{3}}+\frac{c_2}{\sqrt{6}}-\frac{c_3}{\sqrt{10}}$      &  $N_7+N_8$  &  $M_{5}$  &     $M_{5}\overline{d^c} + (M_{5}+N_7+N_8)\overline{L}$      
                        \\
                        $\Sigma_{5_{6,\pm (t_3+t_5)}}$ &  $\frac{20 \sqrt{3} c_1-5 \sqrt{6} c_2+9 \sqrt{10} c_3}{60} $  &  $N_7+N_9$  &  $M_{6}$  &     $M_{6}\overline{d^c} + (M_{6}+N_7+N_9)\overline{L}$      
                        \\
                        $\Sigma_{5_{7,\pm (t_4+t_5)}}$ &           $\frac{5 \sqrt{6} c_2+3 \sqrt{10} c_3}{20}$            &  $N_8+N_9$  &  $M_{7}$  &     $M_{7}\overline{d^c} + (M_{7}+N_8+N_9)\overline{L}$      
                        \\ \hline\hline
        \end{tabular}}
\caption{\small{ Matter curves along with their $U(1)'$ charges, flux data and the corresponding SM content. Note that $N=N_{7}+N_{8}+N_{9}$.}}
\label{Spectable}
\end{table}

There are also singlet fields defined in ~(\ref{Treps}) which play an important r\^ole in the construction of realistic F-theory models. In the present framework,
these  singlet states are parameterised by the vanishing combination
$\pm{(t_{i}-t_{j})}=0\;,\quad{i\neq{j}}$,  therefore, due to $Z_{2}$ monodromy we end up with twelve singlets, denoted by $\theta_{ij}$.
Their $U(1)^{\prime}$ charges and multiplicties are collectively presented in Table \ref{tab:singlets}. Details on their r\^ole in the effective theory will be given in the subsequent sectors.

\begin{table}[H]
        \small
        \centering
        \begin{tabular}{cccc}
                \hline\hline
        Singlet Fields & Weights   & $Q^\prime_{ij}$ $(Q^\prime_{ji})$    & Multiplicity  \\ \hline
        $\theta_{13}$, $(\theta_{31})$ & $\pm(t_1-t_3)$ & $\pm\frac{\sqrt{3}c_1}{2}$   & $M_{13}$, $(M_{31})$\\
        $\theta_{14}$, $(\theta_{41})$ & $\pm(t_1-t_4)$ & $\pm\frac{c_1+2 \sqrt{2} c_2}{2 \sqrt{3}}$  & $M_{14}, (M_{41})$ \\
        $\theta_{15}$, $(\theta_{51})$ & $\pm(t_1-t_5)$ & $\pm\frac{1}{12} \left(2 \sqrt{3} c_1+\sqrt{6} c_2+3 \sqrt{10} c_3\right)$  & $M_{15}$, $(M_{51})$ \\
        $\theta_{34}$, $(\theta_{43})$ & $\pm(t_3-t_4)$ & $\pm\frac{\sqrt{2} c_2-c_1}{\sqrt{3}}    $  & $M_{34}$, $(M_{43})$ \\
        $\theta_{35}$, $(\theta_{53})$ & $\pm(t_3-t_5)$ & $\pm\frac{1}{12} \left(-4 \sqrt{3} c_1+\sqrt{6} c_2+3 \sqrt{10} c_3\right)$ & $M_{35}$, $(M_{53})$ \\
        $\theta_{45}$, $(\theta_{54})$ & $\pm(t_4-t_5)$ & $\pm\frac{1}{4} \left(\sqrt{10} c_3-\sqrt{6} c_2\right)                        $  & $M_{45}$, $(M_{54})$ \\\hline\hline

        \end{tabular}
        \caption{ \small{Singlet fields $\theta_{ij}$  along with their corresponding $U(1)'$ charges  and multiplicities $M_{ij}$. The "$(-)$" sign on the weights and charges refers to the singlets in the parentheses.}}
        \label{tab:singlets}
        \end{table}

\subsection{Anomaly cancellation conditions}

In the previous sections we elaborated on the details of the F-$SU(5)$ GUT  supplemented by a flavour-dependent  $U(1)'$ extension
where this abelian factor is embedded in the $SU(5)_{\perp}\supset E_8$.   Since the effective theory has to be renormalisable 
and  ultra-violet  complete, the  $U(1)'$ extension must be anomaly free. This  requirement imposes significant restrictions
on the $U(1)'$ charges of the spectrum and  consequently, on the coefficients $c_i$ defining the linear combination in~(\ref{Qprime}).
In this section we will work out the anomaly cancellation conditions  to  determine the appropriate linear combinations ~(\ref{Qprime}).
This procedure will also specify  all the possibly allowed  $U(1)'$  charge assignments of the zero-mode spectrum. Consequently, 
each such set of charges will correspond to a distinct low energy model which can give definite predictions to be confronted 
with  experimental data. 
\vspace{1cm}


Although the well known MSSM anomaly cancellation conditions coincide with the chirality condition \eqref{chiralitycondition} imposed by the fluxes,
 there are additional contributions to gauge anomalies due to the extra $U(1)'$ factor. In order to   consistently incorporate
 the new abelian factor into the effective theory, the following six anomaly conditions should be considered:

\ba 
{\cal A}_{331}&:& SU(3)_C SU(3)_C U(1)'\label{331}
\\
{\cal A}_{211}&:&SU(2)_L SU(2)_L U(1)'
\\
{\cal A}_{YY1}&:& U(1)_{Y}U(1)_{Y}U(1)'
\\
{\cal A}_{Y11}&:&U(1)_{Y}U(1)'U(1)'
\\
{\cal A}_{111}&:&U(1)'U(1)'U(1)'
\\
{\cal A}_{G}&:&{\rm Gauge\, Gravity\, Anomaly}\; . \label{GGA}
\ea 
Using the data of Table \ref{Spectable}, it is straightforward to compute the anomaly conditions (\ref{331}-\ref{GGA}). Analytical expressions are given in Appendix \ref{appA}.  
It turns out (up to overall factors) that $\mathcal{A}_{221}=\mathcal{A}_{331}=\mathcal{A}_{YY1}\equiv\mathcal{A}$, where $\mathcal{A}$ depends on $M_i$, $m_{j}$, $N_k$ and linearly on $c_{1,2,3}$. On the other hand, the mixed $\mathcal{A}_{Y11}$ anomaly is not linear on $c_{1,2,3}$ and depends only on the hyperflux integers $N_k$.

The cubic ($\mathcal{A}_{111}$) and gravitational ($\mathcal{A}_{G}$) anomalies depend only on the $U(1)'$ charges (and  flux integers), hence singlet fields come into play. The last terms of \eqref{A111} and \eqref{Agravity} display the contribution from the singlets. Since $Q_{ij}'=-Q_{ji}'$ as a first approximation, we can assume that the singlets always come in pairs ($M_{ij}=M_{ji}$), ensuring this way that their contribution to the anomalies always vanishes.

\subsection{Solution Strategy}

The anomaly conditions displayed above are complicated functions of the $c_{i}$-coefficients and the flux integers $m_{i}$, $M_{j}$ and $N_k$. In order to solve for the $c_{i}$'s we have to deal with the flux integers first. The precise determination of the spectrum in the present construction, depends on the choice of these flux parameters. While there is a relative freedom on the choice and the distribution of generations on the various matter curves, some phenomenological requirements may guide our choices. For example, the requirement for a tree-level top Yukawa coupling suggests that the top quark must be placed on the $10_{1}$ matter curve (see Table~\ref{Spectable})
and the MSSM up-Higgs doublet at $5_1$ since, due to $Z_2$ monodromy, the only renormalisable top-like operator is :
$10_{t_1}10_{t_1}5_{-2t_1}\equiv 10_{1}10_{1}5_{1} $. This suggests the following conditions on some of the flux integers: 
\be
\label{topcondition}
m_{1}={1},\; m_{1}+N\geq{1},\; M_{1}+N\geq{1} .
\ee
Furthermore, a solution to the doublet-triplet splitting problem implies that

\be \label{doubtripcondition}
|N_7|+|N_8|+|N_9|\neq{0}.
\ee   
   
  Additional conditions can be imposed by demanding  certain  properties  of the effective model and a specific zero-mode spectrum. In what follows, we will split our search into two major directions. Namely, minimal models which contain only the MSSM spectrum (no exotics), and models with vector-like pairs. 
  
  For each case we put conditions on the fluxes and then we scan for all possible combinations of flux integers  satisfying all the constraints. Next, each set of flux solutions is applied to the anomaly conditions  \eqref{A331}-\eqref{A111} and we check whether a solution for the $c_i$'s exists. Each solution for the $c_i$'s must also fulfill  the normalisation condition \eqref{norm_condition}.
  
  \section{Models with MSSM spectrum}\label{sec4}
  
  We start with the minimal scenario where the models we are interested in  have the MSSM spectrum accompanied only by pairs of conjugate singlet fields. In particular,  three chiral families  of quarks and leptons of the  MSSM spectrum are ensured by the chirality condition \eqref{chiralitycondition}.

  On top of the conditions \eqref{topcondition} and \eqref{doubtripcondition} we also assume that 
  
  \be\label{Hu_condition}
M_{1}=0,\;N=1  ~,
  \ee 
 
\noindent avoiding this way exotics since $H_{u}$ will be the only MSSM state in $5_1$ matter curve. In addition, absence of exotics  necessarily implies that 

\be\label{noexoticscondition}
m_{i}\geq{0}\; , \; -M_{j}\geq{0}~. 
\ee

Then we search the flux parameter space for combinations of $m_i$, $M_j$ and $N_k$ which respect the conditions  \eqref{chiralitycondition}, \eqref{topcondition}, \eqref{doubtripcondition}, \eqref{Hu_condition} and \eqref{noexoticscondition}. We allow the flux parameters to vary in the range $[-3,3]$.

Our scan identifies fifty-four  sets of flux integers that are consistent with all the MSSM spectrum criteria and a tree-level top term. From these fifty-four flux solutions, only six of them 
yield  a solution for the $c_i$ coefficients with equal pairs of singlets, $M_{ij}=M_{ji}$. This class of solutions are shown in Table \ref{tab:classA_fluxes} and the spectrum of the corresponding models are presented in Table \ref{tab:classA_models}. We refer to this class of models as \emph{Class A}.

\begin{table}[H]
        \small
        \centering
        \begin{tabular}{c|cccc|ccccccc|ccc|ccc}
                \hline\hline
        Model & $m_1$ & $m_2$ & $m_3$ & $m_4$ & $M_1$                                                                             & $M_2$ & $M_3$ & $M_4$ & $M_5$ & $M_6$ & $M_7$ & $N_7$ & $N_8$ & $N_9$                                                                                                                 & $c_1$ & $c_2$ & $c_3$     \\\hline
\textbf{A1} & 1 & 2 & 0 & 0 & 0 & -1 & 0 & 0 & -1 & -1 & 0 & 1 & 0 & 0 & 0 & $ -\frac{1}{2}\sqrt{\frac{3}{2}}$ & $ \frac{1}{2}\sqrt{\frac{5}{2}}$\\
\textbf{A2} & 1 & 0 & 2 & 0 & 0 & 0 & -1 & 0 & -1 & 0 & -1 & 0 & 1 & 0 &$\frac{1}{\sqrt{3}}$ & $ -\frac{1}{2\sqrt{6}}$ & $ -\frac{1}{2}\sqrt{\frac{5}{2}}$\\
\textbf{A3} & 1 & 0 & 0 & 2 & 0 & 0 & 0 & -1 & 0 & -1 & -1 & 0 & 0 & 1 & $\frac{1}{\sqrt{3}}$ & $ -\sqrt{\frac{2}{3}}$ & $ 0$
\\
\textbf{A4} & 1 & 0 & 0 & 2 & 0 & 0 & 0 & 0 & -1 & -1 & -1 & 0 & 0 & 1 & $\frac{1}{\sqrt{3}}$ & $ -\sqrt{\frac{2}{3}}$ & $ 0$\\ 
\textbf{A5} & 1 & 0 & 2 & 0 & 0 & 0 & 0 & 0 & -1 & -1 & -1 & 0 & 1 & 0  &$\frac{1}{\sqrt{3}}$ & $ -\frac{1}{2\sqrt{6}}$ & $ -\frac{1}{2}\sqrt{\frac{5}{2}}$\\
\textbf{A6} & 1 & 2 & 0 & 0 & 0 & 0 & 0 & 0 & -1 & -1 & -1 & 1 & 0 & 0 & 0 & $ -\frac{1}{2}\sqrt{\frac{3}{2}}$ & $ \frac{1}{2}\sqrt{\frac{5}{2}}$\\\hline\hline
        \end{tabular}
        \caption{ \small{MSSM flux solutions along with the resulting $c_{i}$ 's. For this class of models (Class A), singlets come in pairs ($M_{ij}=M_{ji}$).}}
        \label{tab:classA_fluxes}
        \end{table}

\begin{table}[H]
\resizebox{\textwidth}{!}{%
\begin{tabular}{cc||cc||cc||cc||cc||cc}\hline\hline
\multicolumn{2}{c}{Model \textbf{A1}} & \multicolumn{2}{c}{Model \textbf{A2}} & \multicolumn{2}{c}{Model \textbf{A3}} & \multicolumn{2}{c}{Model \textbf{A4}} & \multicolumn{2}{c}{Model \textbf{A5}} & \multicolumn{2}{c}{Model \textbf{A6}} \\\hline
$Q'$   & SM                 & $Q'$   & SM                 & $Q'$   & SM                 & $Q'$   & SM                 & $Q'$   & SM                 & $Q'$   & SM                 \\\hline
0      & $Q+2u^c$           & 0      & $Q+2u^c$           & 0      & $Q+2u^c$           & 0      & $Q+2u^c$           & 0      & $Q+2u^c$           & 0      & $Q+2u^c$           \\
0      & $2Q+u^{c}+3e^{c}$  & -1/2   & -                  & -1/2   & -                  & -1/2   & -                  & -1/2   & -                  & 0      & $2Q+u^{c}+3e^{c}$  \\
1/2    & -                  & 0      & $2Q+u^{c}+3e^{c}$  & 1/2    & -                  & 1/2    & -                  & 0      & $2Q+u^{c}+3e^{c}$  & 1/2    & -                  \\
-1/2   & -                  & 1/2    & -                  & 0      & $2Q+u^{c}+3e^{c}$  & 0      & $2Q+u^{c}+3e^{c}$  & 1/2    & -                  & -1/2   & -                  \\\hline
0      & $H_u$              & 0      & $H_u$              & 0      & $H_u$              & 0      & $H_u$              & 0      & $H_u$              & 0      & $H_u$              \\
0      & $d^{c}+2L$         & -1/2   & $L$                & -1/2   & $L$                & -1/2   & $L$                & -1/2   & $L$                & 0      & $L$                \\
1/2    & $L$                & 0      & $d^{c}+2L$         & 1/2    & $L$                & 1/2    & $L$                & 0      & $L$                & 1/2    & $L$                \\
-1/2   & $L$                & 1/2    & $L$                & 0      & $d^{c}+2L$         & 0      & $L$                & 1/2    & $L$                & -1/2   & $L$                \\
1/2    & $d^c$              & -1/2   & $d^c$              & 0      & -                  & 0      & $d^{c}+L$          & -1/2   & $d^{c}$            & 1/2    & $d^{c}$            \\
-1/2   & $d^c$              & 0      & -                  & -1/2   & $d^c$              & -1/2   & $d^c$              & 0      & $d^{c}+L$          & -1/2   & $d^{c}$            \\
0      & -                  & 1/2    & $d^{c}$            & 1/2    & $d^c$              & 1/2    & $d^c$              & 1/2    & $d^{c}$            & 0      & $d^{c}+L$       \\\hline\hline  
\end{tabular}%
}
\caption{\small{Models with MSSM spectrum plus pairs of singlet fields ($M_{ij}=M_{ji}$).}}
\label{tab:classA_models}
\end{table}

\noindent Note that the SM states of all the models above carry the same charges under the extra $U(1)'$ and differ only on how the SM states are distributed among the various matter curves. In all cases we expect similar low energy phenomenological implications.  

Solutions for the remaining forty-eight set of fluxes arise if we relax the condition $M_{ij}=M_{ji}$ and allow for general multiplicities for the singlets. Scanning  the parameter space, three new classes (named as \emph{Class B}, \emph{Class C} and \emph{Class D}),  of consistent solutions emerge. Some representative solutions from each class\footnote{Each class consists of various flux and $c_{i}$ solutions that results to the same $Q'$ charges. The various models inside a class are differ on how the SM fields distributed on the matter curves.} are shown in Table \ref{tab:mssmfluxgeneral} while the corresponding models are presented in Table \ref{tab:mssm_789}. A complete list of all the flux solutions, the corresponding charges and singlet spectrum is given in Appendix \ref{AppB}.

\begin{table}[H]
        \small
        \centering
        \begin{tabular}{c|cccc|ccccccc|ccc|ccc}
                \hline\hline
        Model & $m_1$ & $m_2$ & $m_3$ & $m_4$ & $M_1$                                                                             & $M_2$ & $M_3$ & $M_4$ & $M_5$ & $M_6$ & $M_7$ & $N_7$ & $N_8$ & $N_9$                                                                                                                 & $c_1$ & $c_2$ & $c_3$     \\\hline
\textbf{B7}& 1 & 0 & 1 & 1 & 0 & -1 & 0 & 0 & -1 & 0 & -1 & 0 & 1 & 0 & $-\frac{\sqrt{5}}{3}$ & $\frac{1}{6}\sqrt{\frac{5}{2}}$  & $ -\frac{1}{2}\sqrt{\frac{3}{2}}$ \\
\textbf{C8}& 1 & 0 & 0 & 2 & 0 & 0 & -1 & 0 & 0 & -1 & -1 & 0 & 0 & 1 & $-\frac{\sqrt{5}}{6}$ & $\frac{7}{12}\sqrt{\frac{5}{2}}$  & $ -\frac{1}{4\sqrt{6}}$ \\
\textbf{D9}& 1 & 1 & 0 & 1 & 0 & 0& 0 & 0 & -1 & -1 & -1 & 0 & 0 & 1 & $\frac{1}{2}\sqrt{\frac{5}{6}}$ & $\frac{5}{8}\sqrt{\frac{5}{3}}$  & $- \frac{3}{8}$ \\\hline\hline
        \end{tabular}
        \caption{ \small{MSSM flux solutions along with the corresponding $c_{i}$ 's for a general singlet spectrum. }}
        \label{tab:mssmfluxgeneral}
        \end{table}

\begin{table}[H]
\resizebox{\textwidth}{!}{%
\begin{tabular}{c||cc||cc||cc}\hline\hline
\multirow{2}{*}{\begin{tabular}[c]{@{Curve}c@{}}\\ \end{tabular}} & \multicolumn{2}{c||}{\underline{\underline{\qquad{Model \textbf{B7}}\qquad}}}     & \multicolumn{2}{c||}{\underline{\underline{\qquad{Model \textbf{C8}}\qquad}}}       & \multicolumn{2}{c}{\underline{\underline{\qquad{Model \textbf{D9}}\qquad}}}     \\
                                                                         & $\sqrt{15}Q'$ & SM              & $\sqrt{15}Q'$ & SM                & $\sqrt{10}Q'$ & SM              \\\hline
$10_{1}$                                                                 & -1             & $Q+2u^c$        & 1/4           & $Q+2u^c$          & 3/4          & $Q+2u^c$        \\
$10_{2}$                                                                 & 3/2          & -               & 3/2           & -                 & -1/2           & $Q+u^{c}+e^{c}$ \\
$10_{3}$                                                                 & -1             & $Q+2e^{c}$      & -9/4          & -                 & -7/4           & -               \\
$10_{4}$                                                                 & 3/2          & $Q+u^{c}+e^{c}$ & 1/4           & $2Q+u^{c}+3e^{c}$ & 3/4          & $Q+2e^{c}$      \\\hline
$5_{1}$                                                                  & 2            & $H_u$           & -1/2          & $H_u$             & -3/2           & $H_u$           \\
$\bar{5}_{2}$                                                            & 1/2          & $d^{c}+2L$      & 7/4           & $L$               & 1/4          & $L$             \\
$\bar{5}_{3}$                                                            & -2             & $L$             & -2            & $d^{c}+2L$        & -1             & $L$             \\
$\bar{5}_{4}$                                                            & 1/2          & $L$             & 1/2           & $L$               & 3/2          & $L$             \\
$\bar{5}_{5}$                                                            & 1/2          & $d^c$           & -3/4          & -                 & -9/4           & $d^{c}+L$       \\
$\bar{5}_{6}$                                                            & 3            & $d^c$           & 7/4           & $d^c$             & 1/4          & $d^{c}$         \\
$\bar{5}_{7}$                                                            & 1/2          & -               & -2            & $d^c$             & -1             & $d^{c}$    \\\hline\hline    
\end{tabular}%
}
\caption{\small{MSSM like models accompanied by a general singlet spectrum.}}
\label{tab:mssm_789}
\end{table}

It is being observed   that for all the models presented so far,  one  of the tenplets $10_2$, $10_{3}, 10_4$ acquires the same $U(1)^{\prime}$ charge
with the $10_1$ matter curve accommodating the top-quark. Thus, at least one of the lightest left-handed quarks will have the same $Q'$ charge with the top quark.  In this case, the corresponding flavour processes associated with these two families are expected to be suppressed. 

Next, we will investigate some  phenomenological aspects of the models presented so far. We first   write down all the possible $SU(5)\times U(1)' $ invariant tree-level Yukawa terms:

$\bullet$ Renormalisable top-Yukawa type operator:
\be
10_{1}10_{1}\bar{5}_{1}~,
\ee

\noindent  which is the only tree-level top quark operator allowed by the $t_{i}$ weights (see Tables~\ref{tab:mssmfluxgeneral},\ref{tab:mssm_789}) thanks to the $Z_2$ monodromy.

$\bullet$ Renormalisable bottom-type quarks operators:

\begin{equation}
10_{1}\bar{5}_{2}\bar{5}_{7},\; 10_{1}\bar{5}_{3}\bar{5}_{6},\; 10_{1}\bar{5}_{4}\bar{5}_{5},\; 10_{2}\bar{5}_{3}\bar{5}_{4},\; 10_{3}\bar{5}_{2}\bar{5}_{4},\; 10_{4}\bar{5}_{2}\bar{5}_{3} \cdot
\end{equation}

\noindent Depending on how the SM states are distributed among the various matters curves, tree level bottom and/or R-parity violation (RPV) terms may exist in the models.  
\subsection{Phenomenological Analysis }
Up till now we 
have sorted out a small number of phenomenologically viable models
 distinguished by their low energy predictions. 
In the remaining of this section, we will focus on Model D9.  The implications of the remaining
 models will be explored in the Appendix. 

Details for the fermion sectors of this model are given in Table \ref{tab:mssm_789}, while the properties of the singlet sector can be found in Appendix \ref{AppB}. In order to achieve realistic fermion hierarchies, we assume the following distribution of the MSSM spectrum  in to the various matter curves:
\begin{equation*}
10_{1}\longrightarrow{Q_{3}+u_{2,3}^{c}},\;\; 10_{2}\longrightarrow{Q_{1}+u_{1}^{c}+e_{1}^{c}},\;\; 10_{4}\longrightarrow{Q_{2}+e_{2,3}^{c}}\;,
\end{equation*}
\begin{equation*}
5_{1}\longrightarrow{H_{u}},\;\; \bar{5}_{2}\longrightarrow{H_{d}},\;\; \bar{5}_{3}\longrightarrow{L_{3}},\;\;\bar{5}_{4}\longrightarrow{L_{2}},\;\;\bar{5}_{5}\longrightarrow{d_{1}^{c}+L_{1}},\;\;\bar{5}_{6}\longrightarrow{d^{c}_{2}},\;\;\bar{5}_{7}\longrightarrow{d^{c}_{3}}\;,
\end{equation*}

\noindent where the indices (1,2,3) on the SM states denote generation.

\noindent\textbf{Top Sector}

The dominant contributions to the up-type quarks descend from the following superpotential terms

\begin{equation}
\begin{split}
W &\supset y_{t}10_{1}10_{1}5_{1}+\frac{y_{1}}{\Lambda}10_{1}10_{2}5_{1}\theta_{13}+\frac{y_{2}}{\Lambda}10_{1}10_{4}5_{1}\theta_{15}+\frac{y_{3}}{\Lambda^{2}}10_{2}10_{4}5_{1}\theta_{13}\theta_{15}\\
&+\frac{y_{4}}{\Lambda^{2}}10_{2}10_{2}5_{1}\theta_{13}^{2}+\frac{y_{5}}{\Lambda^{2}}10_{1}10_{2}5_{1}\theta_{15}\theta_{53}+\frac{y_{6}}{\Lambda^{3}}10_{2}10_{2}5_{1}\theta_{15}\theta_{53}\theta_{13}~,
\end{split}
\end{equation}

\noindent where $y_{i}$'s are coupling constant coefficients and $\Lambda$ is a characteristic  high energy scale of the theory. The operators yield the following mass texture :

\be
M_{u} = v_{u}\left(\begin{array}{ccc}
y_{4}\vartheta_{13}^{2}+y_{6}\vartheta_{15}\vartheta_{53}\vartheta_{13} & y_{3}\vartheta_{13}\vartheta_{15} & y_{1}\vartheta_{13}+y_{5}\vartheta_{15}\vartheta_{53}\\
y_{1}\vartheta_{13}+y_{5}\vartheta_{15}\vartheta_{53} & y_{2}\vartheta_{15} & \varepsilon y_{t}\\
y_{1}\vartheta_{13}+y_{5}\vartheta_{15}\vartheta_{53} & y_{2}\vartheta_{15} & y_{t}\\
\end{array}\right)~,
\ee

\noindent where $v_{u}=\langle{H_{u}}\rangle$, $\vartheta_{ij}=\langle{\theta_{ij}}\rangle/\Lambda$ and $\varepsilon\ll{1}$ is a suppression factor introduced here to capture local effects of Yukawa couplings descending from a common tree-level operator  \cite{Cecotti:2009zf, Aparicio:2011jx, Marchesano:2015dfa}. The matrix has the appropriate structure to explain the hierarchy in the top sector.  

\noindent\textbf{Bottom Sector}

There is one tree-level and several non-renormalisable operators contributing to the down-type quarks. The dominant terms are:

\be
\begin{split}
W &\supset y_{b}10_{1}\bar{5}_{7}\bar{5}_{2}+\frac{\kappa_{1}}{\Lambda}10_{1}\bar{5}_{5}\bar{5}_{2}\theta_{53}+\frac{\kappa_{2}}{\Lambda}10_{1}\bar{5}_{6}\bar{5}_{2}\theta_{43}+\frac{\kappa_{3}}{\Lambda}10_{2}\bar{5}_{7}\bar{5}_{2}\theta_{13}+\frac{\kappa_{4}}{\Lambda^{2}}10_{2}\bar{5}_{6}\bar{5}_{2}\theta_{13}\theta_{43}\\
&+\frac{\kappa_{5}}{\Lambda^{2}}10_{2}\bar{5}_{5}\bar{5}_{2}\theta_{13}\theta_{53}+\frac{\kappa_{6}}{\Lambda^{2}}10_{2}\bar{5}_{7}\bar{5}_{2}\theta_{15}\theta_{53}+\frac{\kappa_{7}}{\Lambda^{3}}10_{2}\bar{5}_{5}\bar{5}_{2}\theta_{15}\theta^{2}_{53}+\frac{\kappa_{8}}{\Lambda^{3}}10_{2}\bar{5}_{6}\bar{5}_{2}\theta_{14}\theta_{43}^{2}+\frac{\kappa_{9}}{\Lambda}10_{4}\bar{5}_{7}\bar{5}_{2}\theta_{15}\\
&+\frac{\kappa_{10}}{\Lambda}10_{4}\bar{5}_{5}\bar{5}_{2}\theta_{13}+\frac{\kappa_{11}}{\Lambda^2}10_{4}\bar{5}_{6}\bar{5}_{2}\theta_{13}\theta_{45}+\frac{\kappa_{12}}{\Lambda^{2}}10_{4}\bar{5}_{5}\bar{5}_{2}\theta_{15}\theta_{53}+\frac{\kappa_{13}}{\Lambda^3}10_{4}\bar{5}_{6}\bar{5}_{2}\theta_{15}\theta_{45}\theta_{53}
\end{split}~,
\ee

\noindent  with $\kappa_{i}$, $y_{b}$
being coupling constant coefficients. These operators generate the following down quark mass matrix:

 \be\label{Wb9}
M_{d} = v_{d}\left(\begin{array}{ccc}
\kappa_{5}\vartheta_{53}\vartheta_{13}+\kappa_{7}\vartheta_{15}\vartheta_{53}^{2} & \kappa_{10}\vartheta_{13}+\kappa_{12}\vartheta_{15}\vartheta_{53} & \kappa_{1}\vartheta_{53} \\
\kappa_{4}\vartheta_{13}\vartheta_{43}+\kappa_{8}\vartheta_{14}\vartheta_{43}^{2}& \kappa_{11}\vartheta_{13}\vartheta_{45}+\kappa_{13}\vartheta_{15}\vartheta_{45}\vartheta_{53} & \kappa_{2}\vartheta_{43}\\
\kappa_{3}\vartheta_{13}+\kappa_{6}\vartheta_{15}\vartheta_{53} & \kappa_{9}\vartheta_{15} & y_{b}\\
\end{array}\right)\; ,
\ee

\noindent where $v_{d}=\langle{H_d}\rangle$ is the VEV of the down-type MSSM Higgs. This matrix is subject to corrections from higher order terms and due to the many contributing operators,   we expect large mixing effects.
 
\noindent\textbf{Charged Lepton Sector}
 
 In the present construction, when flux  pierces  the various matter curves,  the SM generations are distributed on different matter curves. As a consequence, in general, down type quarks and charged lepton sectors emerge from different couplings.

In the present model the common operators between bottom and charged lepton sector are those given in \eqref{Wb9} with couplings $\kappa_{5}$, $\kappa_{7}$, $\kappa_{10}$ and $\kappa_{12}$. All the other contributions descend from the operators

\be 
W \supset 
y_{\tau}10_{4}\bar{5}_{3}\bar{5}_{2}+\frac{\lambda_{1}}{\Lambda}10_{2}\bar{5}_{4}\bar{5}_{2}\theta_{43}+\frac{\lambda_{2}}{\Lambda}10_{2}\bar{5}_{3}\bar{5}_{2}\theta_{53}+\frac{\lambda_{3}}{\Lambda}10_{4}\bar{5}_{4}\bar{5}_{2}\theta_{45}~,
\ee

\noindent where $y_{\tau}$ is a tree level Yukawa  coefficient, $\lambda_{i}$ coupling constants and $\eta\ll{1}$ encodes local tree-level Yukawa coupling effects. Collectively we have the following mass texture for the charged leptons of the model:

\be 
M_{e}=v_{d}\left(\begin{array}{ccc}
\kappa_{5}\vartheta_{53}\vartheta_{13}+\kappa_{7}\vartheta_{15}\vartheta_{53}^{2} & \lambda_{1}\vartheta_{43} & \lambda_{2}\vartheta_{53} \\
 \kappa_{10}\vartheta_{13}+\kappa_{12}\vartheta_{15}\vartheta_{53} & \lambda_{3}\vartheta_{45} & \eta y_{\tau}\\
 \kappa_{10}\vartheta_{13}+\kappa_{12}\vartheta_{15}\vartheta_{53} & \lambda_{3}\vartheta_{45} &y_{\tau}\\
\end{array}\right)\; .
\ee

\noindent\textbf{The $\mu$-\textbf{term}}

\noindent The bilinear term $5_{1}\bar{5}_{2}$ is not invariant under the extra $U(1)'$ symmetry. However, the $\mu$-term appears dynamically through the renormalisable operator:

\be  
\kappa 5_{1}\bar{5}_{3}\theta_{13}\longrightarrow\kappa\langle\theta_{13}\rangle H_{u}H_{d}\equiv{\mu H_{u}H_{d}}\;.
\ee 

\noindent There are no constraints imposed on the VEV of 
singlet field $\theta_{13}$, thus, a  proper tuning of the values of $\kappa$ and  $\langle\theta_{13}\rangle$ can lead to an acceptable $\mu$-parameter, $\mu\sim{\mathcal{O}(TeV)}$. As a result, the $\theta_{13}$ singlet which also contributes to the quarks and charged lepton sectors, must  receive VEV at some  energy scale  close to the TeV region.

We also note that some of the singlet fields   couple to the left-handed neutrinos and, in principle, can play 
the r\^ole of their right-handed partners. In particular, as suggested in~\cite{Beasley:2008kw}, the six-dimensional massive KK-modes
which correspond to the  neutral singlets identified by the $Z_2$ symmetry $\theta_{12}\equiv \theta_{21}$ are the most appropriate
fields to be identified as $\theta_{12}\to \nu^c$ and $\theta_{21}\to \bar\nu ^c$ so that a Majorana mass term $M_N \nu^c \bar\nu^c$
is possible. We will not elaborate on this issue any further; some related phenomenological analysis can be found in~\cite{Antoniadis:2013joa}.

\noindent\textbf{CKM matrix}

The square of the fermion mass matrices obtained so far can be diagonalised via the unitary matrices $V_{f_{L}}$. The various coupling constants and VEVs can be fitted to make the diagonal mass matrices satisfy the appropriate mass relations at the GUT scale. In our analysis we use the RGE results  for a large $\tan\beta=v_{u}/v_{d}$ scenario produced in Ref.~\cite{Ross:2007az}. In addition, the combination $V_{u_{L}}V_{d_{L}}^{\dagger}$ must  resemble as close as possible the CKM matrix. 

For the various parameters of the present model, we use a natural set 
of  numerical values  
$$\kappa_{i}\simeq{1},\; y_{1}=y_{4}=y_{5}=y_{6}=25y_{2}=25y_{3}\simeq{0.5},\; \epsilon=10^{-4},\; y_{t}=0.5,\; y_{b}=0.36 \; .
$$
Then, the singlet VEV's $\vartheta_{ij}$ are  fitted to:
\[ \vartheta_{13}\simeq{3.16\times{10^{-12}}}, \vartheta_{14}\simeq{3.98\times{10^{-3}}},\; \vartheta_{15}\simeq{10^{-1}},\; \vartheta_{43}\simeq{1.9\times{10^{-2}}},\; \vartheta_{53}\simeq{6.94\times{10^{-3}}},\;  \vartheta_{45}\simeq{10^{-2}}~. \]
\noindent 
For the up and down quark diagonalising matrices, they yield
\be
V_{u_{L}}=\left(
\begin{array}{ccc}
 -1 & -0.000694 & 0.000694 \\
 0.000694 & -1 & 0.000116 \\
 0.0006939 & 0.000116 & 1 \\
\end{array}
\right),\;
V_{d_{L}}=\left(
\begin{array}{ccc}
 -0.9738 & 0.2273 & 0.00674 \\
 -0.2266 & -0.9726 & 0.0519 \\
 0.0183 & 0.04908 & 0.9986 \\
\end{array}
\right)~\cdot 
\ee

\noindent The resulting CKM matrix is in agreement with 
the experimentally measured values
\be
|V_{CKM}|\simeq\left(
\begin{array}{ccc}
 0.973659 & 0.227932 & 0.00601329 \\
 0.227325 & 0.972437 & 0.0518632 \\
 0.0176688 & 0.04913 & 0.998636 \\
\end{array}
\right).
\ee
\noindent It is clear that the CKM matrix is mostly influenced  by the bottom sector while $V_{u_{L}}$ is almost diagonal and unimodular.

Next, we compute the unitary matrix $V_{e_{L}}$ which diagonalises the charged lepton mass matrix. The correct Yukawa relations and the
charged lepton mass spectrum are obtained  for 

\be
V_{e_{L}}=\left(
\begin{array}{ccc}
 -0.801463 & 0.597943 & 0.0110641 \\
 -0.597877 & -0.801539 & 0.00888511 \\
 0.0141811 & 0.000506117 & 0.999899 \\
\end{array}
\right)~,
\ee

\noindent where the remaining parameters were fitted to: $\lambda_{1}=0.4, \lambda_{2}=\lambda_{3}=1, \eta=10^{-4}$ and $y_{\tau}=\simeq{0.51}$.

\noindent\textbf{R-parity violating terms}

In the model under discussion,  several tree-level  as well as bilinear operators leading to RPV effects remain invariant under all the symmetries of the theory. More precisely, the tree-level operators :
\ba 
&10_{1}\bar{5}_{3}\bar{5}_{6}\longrightarrow\lambda^{\prime}Q_{3}L_{3}d_{2}^{c}\,,\\
&10_{2}\bar{5}_{3}\bar{5}_{4}\longrightarrow\lambda L_{3}L_{2}e_{1}^{c}\,, 
\ea
\noindent violate both lepton and baryon number. Notice however, the absence of $u^{c}u^{c}d^{c}$ type of RPV terms which in combination with $QLd^c$ terms can spoil the stability of the proton.

\noindent
There  also exist bilinear RPV terms descending from tree-level operators. In the present model, these are:
\be
5_{1}\bar{5}_{3}\theta_{14}\; ,\;\; \; 5_{1}\bar{5}_{4}\theta_{15}\; .
\ee
\noindent The effect of these terms strongly depend on the dynamics of the singlets, however it would be desirable to completely eliminate such operators. 

One can impose an R-symmetry by hand  \cite{Dudas:2010zb} or to investigate the geometric origin of discrete $Z_{N}$ symmetries that can eliminate such operators  \cite{Antoniadis:2012yk}-\cite{Romao:2015jrh}. In addition, the study of such Yukawa coefficients  at a local-level, shows that they can be suppressed for wide regions of the flux parameter space \cite{CrispimRomao:2016tww}. Since in this work we focus mostly in $Z^{\prime}$ flavour changing effects\footnote{Notice however that some RPV terms of the type $Qld^{c}$ and $lle^{c}$ contribute to flavour violation processes, see   \cite{deGouvea:2000cf, Domingo:2018qfg}. For an explanation of the LHCb anomalies through RPV interactions see \cite{Earl:2018snx, Hu:2020yvs, Altmannshofer:2020axr}.}, we will assume that one of the aforementioned mechanisms protects the models from unwanted RPV terms.

\subsection{$Z^{\prime}$ bounds for Model D9}

Having obtained the $V_{f}$ matrices for the top/bottom quark and charged lepton sectors, it is now straightforward to compute the flavour mixing matrices $Q^{\prime}_{f_{L}}$ defined in \eqref{mixingchargematrix}. These matrices, along with the $Z^{\prime}$ mass $(M_{Z^{\prime}})$ and gauge coupling $(g^{\prime})$, enter the computation of the various flavour violating observables described in Section \ref{sec2}. Hence, we can use the  constraints on these observables in order to derive bounds for the $Z^{\prime}$ mass and gauge coupling or, more precisely, for the ratio $g^{\prime}/M_{Z^{\prime}}$. In any case, the so derived bounds must be in accordance with LHC bounds coming from  dilepton and diquark channels \cite{Aaboud:2019roo, Aad:2019fac, CMS:2019tbu}. For heavy $Z^{\prime}$ searches, the LHC bounds on neutral gauge boson masses are strongly model dependent. For  most of the GUT inspired $Z^{\prime}$ models, masses around $\sim{2-3}$ TeV are excluded.

In the model  at hand, we have seen that the lightest generations of the left-handed quarks have different $U(1)^{\prime}$ charges. Consequently,  strong constraints on the $Z^{\prime}$ mass  are expected to come from the $K-\bar{K}$ mixing bounds. Hence, we first start from  the $K-\bar{K}$ system. 

\subsubsection*{$K^{0}-\bar{K}^{0}$ mixing}

Using eq. \eqref{eq:mesonmix} we find for the Kaon oscillation mass split that :

\[\Delta M^{Z^{\prime}}_{K}\simeq{3.967\times{10^{-14}} \left(\frac{g^{\prime}}{M_{Z^{\prime}}}\right)^{2}} \cdot \]

\begin{figure}[t]
\centering
 \includegraphics[width=0.75\linewidth]{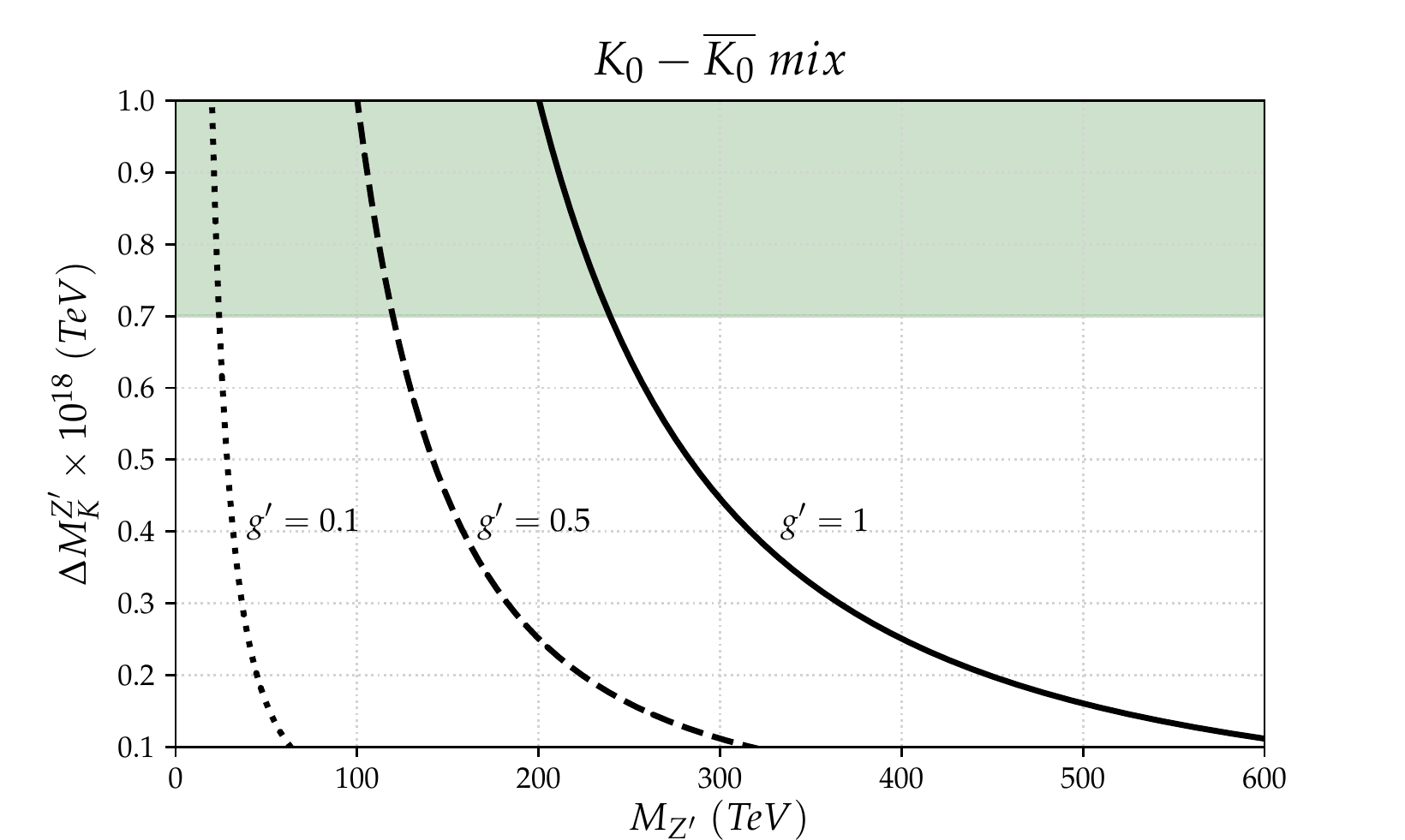}
 \caption{\small{Bounds to the neutral gauge boson mass $M_{Z^{\prime}}$ of Model D9 due to $K_{0}-\bar{K_{0}}$ mixing effects. The vertical axis displays $Z^{\prime}$ contributions ($\Delta M_{K}^{Z^{\prime}}$) to the mass split of the neutral Kaon system. Dotted, dashed and solid black curves correspond to gauge coupling values: $g^{\prime}=0.1$, $0.5$ and $1$ respectively. The shaded region is excluded due the constrain $\Delta M_{K}^{NP}<0.2\Delta M_{K}^{exp}$.}}
 \label{fig:kkmix}
\end{figure}

\noindent The results are plotted  in Figure \ref{fig:kkmix}. As  expected, the Kaon system puts strong bounds on $M_{Z^{\prime}}$. To get an estimate, for $g^{\prime}\simeq{0.5}$ the constraint in \eqref{eq:kaonconstraint} implies that $M_{Z^{\prime}}\gtrsim{120}$ TeV which lies far above the most recent collider searches.

\subsubsection*{$B_{s}^{0}-\bar{B}_{s}^{0}$ mixing}

  From equation \eqref{eq:bbmixwilson} we have that :

\[ C_{bs}^{LL}\approx{1.9\times{10^{-5}}\left(\frac{g^{\prime}\;\mathrm{TeV}}{M_{Z^{\prime}}}\right)^{2}} \]

\noindent which is too small in magnitude to significantly contribute  to $\Delta M_{s}$. This happens because the $U(1)^{\prime}$ charges of $b_{L}$ and $s_{L}$ are equal.

\subsubsection*{$D^{0}-\bar{D}^{0}$ mixing}

 For $M_{D}\simeq{1.86483}$ GeV \cite{Tanabashi:2018oca} and using for the decay constant the value $f_{D}\simeq{212}$ MeV found in \cite{Aoki:2019cca} , the equation \eqref{eq:mesonmix} gives:
\[\Delta M^{Z^{\prime}}_{D}\simeq{2.71\times{10^{-18}} \left(\frac{g^{\prime}\;\;\mathrm{TeV}}{M_{Z^{\prime}}}\right)^{2}} .\]
\noindent Then, for $\Gamma_{D}=1/\tau_{D}\simeq{2.43843}$ $(ps)^{-1}$ \cite{Tanabashi:2018oca} we have that

\[ x_{D}:=\frac{\Delta M_{D}}{\Gamma_{D}}\simeq{0.0017}\left(\frac{g^{\prime}\;\;TeV}{M_{Z^{\prime}}}\right)^2 \]

\noindent which always obeys the bound $x_{D}\leqslant{0.32}$.

\subsubsection*{$P^{0}\rightarrow{l_{i}\bar{l_{i}}}$ decays}

We have found that all the $Z^{\prime}$ contributions are well suppressed when compared to the experimental bounds. 
As an example, consider the decay $B_{d}^{0}\rightarrow{\mu^{+}\mu^{-}}$. Using eq. \eqref{eq:Ptolili} we obtain that

\[ \mathrm{Br}(B_{d}^{0}\rightarrow{\mu^{+}\mu^{-}})\simeq{5.34\times{10^{-9}} \left(\frac{g^{\prime}\;\;\mathrm{TeV}}{M_{Z^{\prime}}}\right)^{4}}\]

\noindent which always satisfies the experimental bound $\mathrm{Br}(B_{d}^{0}\rightarrow{\mu^{+}\mu^{-}})<1.6^{+1.6}_{-1.4}\times{10^{-10}}$, for $g^{\prime}<1$ and $M_{Z^{\prime}}\sim{\mathcal{O}(TeV)}$. Similar results were obtained for lepton flavour violating decays of the form $P^{0}\rightarrow{l_{i}\bar{l_{j}}}$.

\subsubsection*{Muon anomalous magnetic moment and $\mu\rightarrow{e\gamma}$}

Our results imply that $Z^{\prime}$ contributions to $\Delta a_{\mu}$ are always smaller than the observed discrepancy. Even for the limiting case where $g^{\prime}=1$ and $M_{Z^{\prime}}=1$ TeV our computations return: $\Delta a_{\mu}^{Z^{\prime}}\simeq{3\times{10^{-11}}}$. This suggests that for small $Z^{\prime}$ masses the model can explain the observed $(g-2)_{\mu}$ anomaly. However  for larger $M_{Z^{\prime}}$ values implied from the Kaon system the results are very suppressed.

For LFV radiative decays of the form $l_{i}\rightarrow{l_{j}\gamma}$, the strongest bounds are expected from the muon channel. For $g^{\prime}=1$, the present model predicts that $M_{Z^{\prime}}\gtrsim{1.3}$ TeV if the predicted $\mu\rightarrow{e\gamma}$ branching ratio is to satisfy the experimental bounds.  Tau decays ($\tau\rightarrow{e\gamma}$, $\tau\rightarrow{\mu\gamma}$) are well suppressed, due to the short lifetime of the tau lepton.

\begin{figure}[t]
\centering
  \includegraphics[width=0.75\linewidth]{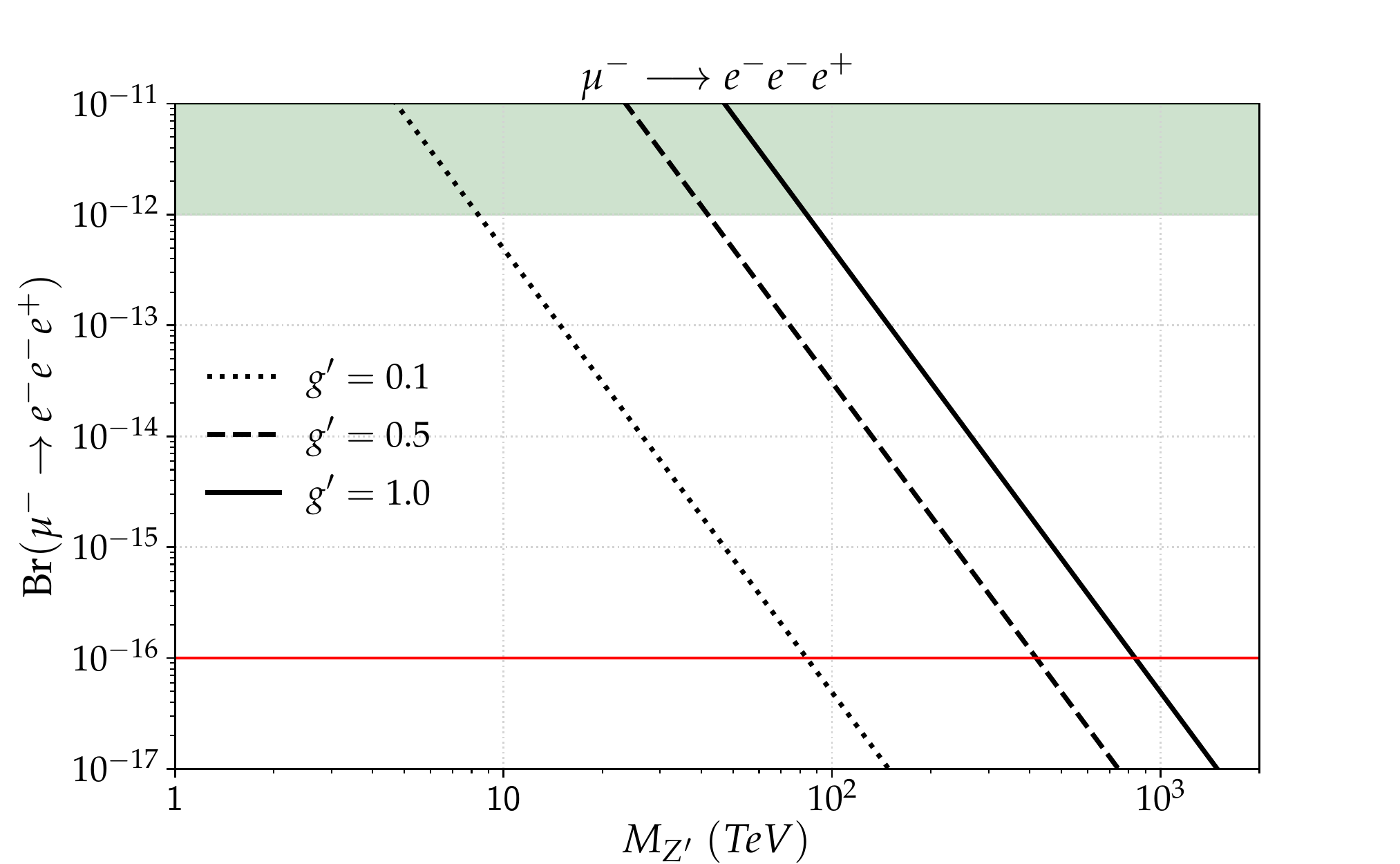}
 \caption{\small{Bounds to the neutral gauge boson mass $M_{Z^{\prime}}$ as predicted in Model D9  from $Z^{\prime}$ contributions to the lepton flavour violation decay $\mu^{-}\rightarrow{e^{-}e^{-}e^{+}}$. The plot shows the branching ratio of the decay as function of the $Z^{\prime}$ mass for various values of the gauge coupling $g^{\prime}$. Both axis are in logarithmic scale. Dotted, dashed and solid black curves correspond to $U(1)^{\prime}$ gauge couplings: $g^{\prime}=0.1$, $0.5$ and $1$ respectively. The shaded region is excluded due to the current experimental bound: $\mathrm{Br}(\mu^{-}\rightarrow{e^{-}e^{-}e^{+}})<10^{-12}$. The red horizontal line represents the estimated reach of future $\mu\rightarrow{3e}$
experiments.}}
 \label{fig:muto3e}
\end{figure}%

\subsubsection*{$\mu^{-}\rightarrow{e^{-}e^{-}e^{+}}$}

While all the three body lepton decays of the form $l_{i}\rightarrow{l_{j}l_{j}\bar{l_{k}}}$ are suppressed for the tau channel, strong constraints are obtained  from the muon decay $\mu^{-}\rightarrow{e^{-}e^{-}e^{+}}$. In particular, the model predicts that

\[ \mathrm{Br}(\mu^{-}\rightarrow{e^{-}e^{-}e^{+}})\simeq{4.92\times{10^{-5}} \left(\frac{g^{\prime}\;\;\mathrm{TeV}}{M_{Z^{\prime}}}\right)^{4}}.\]

\noindent The results are compared with the experimental bounds in Figure \ref{fig:muto3e}. We observe that, for $g^{\prime}={0.5}$ (dashed line in the plot) we receive $M_{Z^{\prime}}\gtrsim{42}$ TeV in order the model to satisfy the current experimental bound, $\mathrm{Br}(\mu^{-}\rightarrow{e^{-}e^{-}e^{+}})<10^{-12}$. While the constraints coming from this decay are stronger than the other lepton flavour violating processes discussed so far, they still are not compatible with the restrictions descending from the Kaon system.

However, important progress is expected by future lepton flavour violation related experiments \cite{Renga:2019mpg}. In particular, the \emph{Mu3e} experiment at PSI \cite{Blondel:2013ia} aim to improve the experimental sensitivity to $\sim{10^{-16}}$. In the absence of a signal, three-body LFV muon decays can then be excluded for $\mathrm{Br}(\mu^{-}\rightarrow{e^{-}e^{-}e^{+}})<10^{-16}$. In Figure \ref{fig:muto3e} the red horizontal line represents the estimated reach of future $\mu\rightarrow{3e}$ experiments. For, $g^{\prime}=0.5$ we find that $M_{Z^{\prime}}\gtrsim{420}$ TeV in order the predicted  branching ratio is to satisfy the foreseen \emph{Mu3e} experimental bounds. Hence, for the present model, the currently dominant bounds from the Kaon system will be exceeded in the near future by the limits of the upcoming $\mu^{-}\rightarrow{e^{-}e^{-}e^{+}}$ experiments.

\subsubsection*{$R_{K}$ anomalies}

The bounds derived  from the Kaon oscillation system and the three-body decay $\mu\rightarrow{e^{-}e^{-}e^{+}}$ leaves no room for a possible explanation of the observed $R_{K}$ anomalies. Indeed, for the relevant Wilson coefficient the model predicts that 
\[ C_{9}\approx{-0.079\left(\frac{g^{\prime}\;\mathrm{TeV}}{M_{Z^{\prime}}}\right)^{2}} \]

\noindent which has the desired sign ($C_{9}<0$), but for $M_{Z^{\prime}}\sim{200}$ TeV and $g^{\prime}\simeq{1}$ the resulting value is too small  to explain the observed B meson anomalies.

Similar phenomenological analysis have been performed for all the other models presented so far. A discussion on their flavour violation bounds is given in Appendix \ref{appC}. Collectively, the results are very similar with those of Model D9. For all the $U(1)^{\prime}$ models with MSSM spectrum the dominant bounds on $M_{Z^{\prime}}$ comes from  $K^{0}-\overline{K^{0}}$ oscillation effects and the muon decay $\mu\rightarrow{e^{-}e^{-}e^{+}}$. 

It is clear from the analysis so far that a successful explanation of the LHCb anomalies  in the present F-theory framework, requires the use of some other type of mechanism. A common approach, is the explanation of the LHCb anomalies through the mixing of the conventional SM matter with extra vector-like fermions  \cite{Allanach:2015gkd}-\cite{Kawamura:2019hxp}. Next, we present such an F-theory model while a full classification of the various F-theory models with a complete family of vector-like fermions will be presented in a future work.

\section{Models with vector-like exotics}\label{sec5}

We expand our analysis to models with the MSSM spectrum + vector-like (VL) states forming complete $(10+\overline{10})$, $(5+\bar{5})$ pairs under the $SU(5)$ GUT symmetry. Hence, as in the previous study, we choose  appropriate  fluxes, solve the anomaly cancellation conditions, and derive the $U(1)^{\prime}$ charges of all the models with additional vector-like families. 

Among the various models, particular attention is paid  to models with 
different $U(1)^{\prime}$ charges for the VL states, while keeping  universal the $U(1)^{\prime}$ charges for the SM fermion families. This way one can explain the observed B-meson anomalies due to the mixing of the SM fermions with the VL exotics while at the same time controlling other flavour violation observables. A model with these properties (first derived in \cite{Romao:2017qnu}) is materialised with the following set of fluxes:
\begin{equation*}
m_{1}=2\;,\;\; m_{2}=m_{3}=-m_{4}=1\;,\;\; M_{1}=M_{2}=M_{3}=M_{7}=0\;,\;\;M_{4}=-M_{6}=1\;,\;\;M_{5}=-3~,
\end{equation*}

\noindent which through anomaly cancellation gives the solution $(c_{1},\;c_{2},\;c_{3})=(\frac{\sqrt{3}}{2},\;-\frac{1}{4}\sqrt{\frac{3}{2}},\;\frac{1}{4}\sqrt{\frac{5}{2}})$. This corresponds to the following $U(1)^{\prime}$ charges for the various matter curves

\begin{align*}
&10_{1}:\frac{1}{4}\;,\;\; 10_{2}:-\frac{1}{2}\;,\;\;10_{3}:\frac{1}{4}\;,\;\;10_{4}:-\frac{1}{4}~,\\
&5_{1}:-\frac{1}{2}\;,\;\;5_{2}:\frac{1}{4}\;,\;\;5_{3}:-\frac{1}{2}\;,\;\;5_{4}:0\;,\;\;5_{5}:\frac{1}{4}\;,\;\;5_{6}:\frac{3}{4}\;,\;\;5_{7}:0\; .
\end{align*}

Assuming the following distribution of the fermion generations and Higgs fields into matter curves
\begin{align*}
10_{1}&\longrightarrow{Q_{2,3}+u_{1,2,3}^{c}}+e_{3}^{c}\;,\;\; 10_{2}\longrightarrow{Q_{4}+u_{4}^{c}+e_{4}^{c}}\;,\;\;10_{3}\longrightarrow{Q_{1}+e_{1,2}^{c}\;,\;\;}\bar{10}_{4}\longrightarrow{\overline{Q_{4}}+\overline{u_{4}^{c}}+\overline{e_{1}^{c}}}\;, \\
5_{1}&\longrightarrow{H_u},\;\; \bar{5}_{2}\longrightarrow{L_{1}},\;\; \bar{5}_{3}\longrightarrow{H_d},\;\;{5}_{4}\longrightarrow{\overline{d_{4}^{c}}},\;\;\bar{5}_{5}\longrightarrow{d_{1,2,3}^{c}+L_{2,3}},\;\;\bar{5}_{6}\longrightarrow{d^{c}_{4}+L_{4}},\;\;{5}_{7}\longrightarrow{\overline{L_{4}}}~,
\end{align*}

\noindent we obtain the desired $U(1)^{\prime}$ charge assignment where all the SM families appear with a common charge ($Q_{1,2,3}^{\prime}=1/4$) while those of the VL states  are non-universal.

Here $Q_{i}, u_{i}^{c}, e^{c}_{i},L_{i},\bar{d}^{c}_{i}$ with $i=1,2,3$ refer to the three SM fermion generations while $u_{4}^{c}$, $\bar{u}_{4}^{c}$, $Q_{4}$, $\overline{Q}_{4}$, $e_{4}$, $\overline{e}^{c}_{4}$, $L_{4}$, $\bar{L}_{4}$, $d_{4}$, $\overline{d}^{c}_{4}$ represent the extra VL states. In a simplified notation, the components of the SM doublets are defined as $Q_{i}=(u_{i},d_{i})$ and similarly for the lepton doublets $L_{i}$. The components of the exotic doublets are $Q_{4}\equiv{(U^{\prime},D^{\prime})}$ and  $\overline{Q}_{4}\equiv{(\bar{D}^{\prime}, \bar{U}^{\prime})}$ and similar for the lepton exotic doublet. For the exotic singlets we use the notation $u_{4}^{c}=\bar{U}$, $\overline{u^{c}}_{4}=U$ and similar $e^{c}_{4}=\bar{E},\; \overline{e_{4}^{c}}=E$, $d^{c}_{4}=\bar{D},\; \overline{d_{4}^{c}}=D$.

The various mass terms can be written in a $5\times{5}$ notation as $\mathrm{F}_{R}\mathbf{M_{F}}\mathrm{F}_{L}$ where $\mathrm{F}_{R}=(f^{c}_{i},\bar{F},\bar{F^{\prime}})$ and $\mathrm{F}_{L}=(f_{i},F^{\prime},F)^{T}$ with $f=u,d,e$ and $F=U,D,E$.
We will focus on the down-type quark sector. The up quark sector can be treated similarly, while the parameters can be adjusted in such a way so that the CKM mixing is ensured. The various invariant operators yielda mass matrix of the form

\be
M_{d}=\left(
\begin{array}{ccccc}
k_0 \vartheta _{14} \vartheta _{54}  v_d &k \varepsilon ^3 \vartheta _{54}  v_d &k \varepsilon ^2 \vartheta _{54}  v_d & k_4 \vartheta _{14} \vartheta _{53}  v_d & k_3 \vartheta _{14} \theta_{53}  \\
 k_{0} \vartheta _{14} \vartheta _{54}  v_d &k \varepsilon ^2 \vartheta _{54} v_d & k\varepsilon  \vartheta _{54}  v_d &k_{4} \vartheta _{14}\vartheta _{53} v_d & k_{3} \vartheta _{14} \theta_{53}  \\
 k_{0}\vartheta_{14} \vartheta_{54} v_d & k \varepsilon  \vartheta_{54} v_d & k\vartheta _{54}  v_d & k_{4}\vartheta _{14} \vartheta _{53}  v_d &k_{3} \vartheta _{14} \theta_{53} \\
 k_{2}\theta _{14} v_d & k_1 \xi  v_d & k_1 v_d & k_{9}\vartheta _{13}  v_d &  k_{10}\theta_{13} \\
  k_6 \theta_{54} v_d &  k_5 \xi \theta_{51}  &  k_5 \theta_{51} &  k_8 \theta_{53} & k_{7}\vartheta _{14} \vartheta _{53}  v_u \\
\end{array}
\right)~,
\ee

\noindent where $k$'s are coupling constant coefficients and $\varepsilon$, $\xi$ are small constant parameters encode local Yukawa effects. Here we represent the singlet VEVs simply as $\theta_{ij}=\langle\theta_{ij}\rangle$ while $\vartheta_{ij}$ represents the ratio $\langle\theta_{ij}\rangle/\Lambda$.  

In order to simplify the matrix we consider that some terms are very small and that approximately vanish. In particular, we assume that $k_{2}=k_{3}=k_{5}\theta_{51}=k_{6}=k_{7}\vartheta_{14}\vartheta_{53}\approx{0}$. Moreover,  we introduce the following simplifications 
\[ k\vartheta_{54}v_{d}=m\;,\; k_{0}\vartheta_{54}\vartheta_{14}v_{d}=\alpha m\;,\;\; k_{4}\vartheta_{14}\vartheta_{53}=\gamma\xi\;,\;\; k_{9}\vartheta_{13}v_{d}=\beta\mu\;,\;\; k_{10}\theta_{13}\simeq{k_{8}}\theta_{53}=M\;,\; \varepsilon\approx{\xi}~,\]
 \noindent where the mass parameter $M$  characterises the VL scale while $m=k\vartheta_{54}v_{d}$  is related to the low energy EW scale. We have also assumed that the small Yukawa parameters are identical $\varepsilon\approx{\xi}$. With these modifications the matrix  takes the following simplified form
 
\be
M_{d}\approx \left(
\begin{array}{ccccc}
 \alpha  m & m \xi ^3 & m \xi ^2 & \gamma  \xi  v_d & 0 \\
 \alpha  m & m \xi ^2 & m \xi  & \gamma  \xi  v_d & 0 \\
 \alpha  m & m \xi  & m & \gamma  \xi  v_d & 0 \\
 0 & k_1 \xi  v_d & k_1 v_d & \beta \mu & M \\
 0 & 0 & 0 &  M & 0 \\
\end{array}
\right)\; .
\ee

\noindent The local Yukawa parameter $\xi$ connects the VL sector with the physics at the EW  scale so we will use this this small parameter to express the mixing between the two sectors. We proceed by perturbatively diagonalizing the down square mass matrix ($M_{d}^{2}$) using $\xi$ as the expansion parameter.

Setting $k_{1}\approx{0}$, $\gamma\;v_{d}=c \mu$ and keeping up to $\mathcal{O}(\xi)$ terms we write the mass square matrix in the form $M^{2}_{d}\approx \mathbf{A}+\xi\;\mathbf{B}$ where:

\begin{align} 
\mathbf{A}=\left(
\begin{array}{ccccc}
 \alpha ^2 m^2 & \alpha ^2 m^2 & \alpha ^2 m^2 & 0 & 0 \\
 \alpha ^2 m^2 & \alpha ^2 m^2 & \alpha ^2 m^2 & 0 & 0 \\
 \alpha ^2 m^2 & \alpha ^2 m^2 & (\alpha ^{2} +1)m^2 & 0 & 0 \\
 0 & 0 & 0 & M^2 & \beta\mu M \\
 0 & 0 & 0 & \beta\mu M & M^2 \\
\end{array}
\right)\;,\;\; \mathbf{B}=\left(
\begin{array}{ccccc}
 0 & 0 & 0 & c \beta \mu^{2} & c\mu  M   \\
 0 & 0 & m^2   & c \beta \mu^{2} & c\mu  M   \\
 0 & m^2  & 0 & c \beta \mu^{2} & c \mu  M   \\
 c \beta \mu^{2} & c \beta \mu^{2} & c \beta \mu^{2} & 0 & 0 \\
 c\mu  M   & c\mu  M   & c\mu  M   & 0 & 0 \\
\end{array}
\right)
\end{align}

\noindent The block-diagonal  matrix $\mathbf{A}$, is the leading order part of the mass square matrix and can be diagonalised by a unitary matrix $V_{b_{L}}^{0}$ as $V_{b_{L}}^{0}\mathbf{A}V_{b_{L}}^{0 T}$ . Its mass square eigenvalues are 
\begin{align}
\nonumber x_{1}&=0\;,\;\; x_{2}=\frac{m^2}{2} \left(1+3\alpha^{2}-\sqrt{1-2\alpha^{2}+9\alpha^4}\right)\;,\;\;x_{3}=\frac{m^2}{2} \left(1+3\alpha^{2}+\sqrt{1-2\alpha^{2}+9\alpha^4}\right)\\
x_{4}&=M(M-\beta \mu)\;,\;\;x_{5}= M(M+\beta\mu)~,
\end{align} 

\noindent where $x_{1,2,3}$ correspond to the mass squares of the three down type quark generations $d_{1,2,3}$ respectively. At this stage we ignore the small mass of the first generation down quark which can be generated by high order corrections. For the second  and third generation we observe that the ratio $\sqrt{x_{2}/x_3}$ depends only on the parameter $\alpha$. Hence from the known ratio $m_{s}/m_{b}$ we estimate that $\alpha\simeq{10^{-2}}$.

The corresponding normalised eigenvectors which form the columns of the diagonalising matrix are
\begin{equation}
\begin{split}
v_{b1}^0= \frac{1}{\sqrt{2}}\left(
\begin{array}{c}
 -1 \\
1\\
0\\
0\\
0\\
\end{array}\right)\;,\; &v_{b2}^0=\frac{1}{\sqrt{1+2q^2}}\left(
\begin{array}{c}
 q \\
q\\
1\\
0\\
0\\
\end{array}\right)\;,\;v_{b3}^0=\frac{-1}{\sqrt{2(1+2q^{2})}}\left(
\begin{array}{c}
_1 \\
1\\
-2q\\
0\\
0\\
\end{array}\right)\;,\; \\
&v_{b4}^0=\frac{1}{\sqrt{2}}\left(
\begin{array}{c}
 0 \\
0\\
0\\
-1\\
1\\
\end{array}\right)\;,\;\; v_{b5}^0=\frac{1}{\sqrt{2}}\left(
\begin{array}{c}
 0 \\
0\\
0\\
1\\
1\\
\end{array}\right)~,
\end{split}
\end{equation}

\noindent where $q=1-\frac{m^2}{x_2}$ depends only on the parameter $\alpha$, since $x_{2}\sim{m^2}$. 

The corrections to the above eigenvectors due to the perturbative part $\xi\mathbf{B}$ are given by the relation

\begin{equation}
v_{b_{i}}\approx v_{b_{i}}^{0}+\xi\sum_{j\neq{i}}^{5}\frac{(V_{b_{L}}^{0}\mathbf{B}V_{b_{L}}^{0\dagger})_{ji}}{x_{i}-x_{j}}  v_{b_{j}}^{0}
\end{equation}

\noindent where the second term displays the $\mathcal{O}(\xi)$ corrections to the basic eigenvectors of the leading order matrix $\mathbf{A}$. The corrected diagonalizing matrices schematically receive the form $V_{b_{L}}=V_{b_{L}}^{0}+\xi V_{b_{L}}^{1}$ and through them the mixing parameter $\xi$ enters on the computation of the various flavour violation observables.

For the explanation of the LHCb anomalies we will consider that perturbative corrections are important for the corresponding $bs$ coupling while almost vanish for the other flavour mixing coefficients. That way , due to the universal $U(1)^{\prime}$ charges of the SM matter most of the flavour violation process are suppressed.  

Assuming that the corresponding lepton contribution is $(Q^{\prime}_{e_{L}})_{22}\approx 1$ and for $\alpha=0.016$ we find for the $b\rightarrow{s}$ transition matrix element that :

\be\label{23vlmix}
(Q^{\prime}_{d_{L}})_{23}\approx Q^{\prime}_{1,2,3}\xi^{2}-0.7 (c\beta)^{2}\left(\frac{m}{M}\right)^{2}\left(\frac{\mu}{M}\right)^{4}Q^{\prime}_{4}\xi^2
\ee

\noindent where $Q^{\prime}_{1,2,3}=1/4$ is the common charge of the MSSM fermions and $Q^{\prime}_{4}=-1/2$ is the charge of the extra matter descending from $10_{2}$ matter curve. Note that the corresponding $U(1)^{\prime}$ charge of the states descending from $5_{4}$ matter curve is zero and consequently does not contribute to the above formula. 

It is clear from equation \eqref{23vlmix} that the first term is dominant since the second one is suppressed due to the large VL mass scale characterized by the parameter $M$. Hence, keeping only the first term we have through equation \eqref{c9wilson} that

\be
C_{9}\approx -963\left(\frac{g^{\prime}}{M_{Z^{\prime}}}\right)^{2} Q^{\prime}_{1,2,3} \xi^{2}
\ee

\noindent and for $g^{\prime}\lesssim{1}$, $M_{Z^{\prime}}\gtrsim{4}$ TeV and $\xi^2\sim{\mathcal{O}(10^{-1})}$ predicts $C_{9}\approx{-1}$ which is  the desired value for the explanation of the LHCb anomalies. It is emphasised here that this approach is valid in the
small $\xi <1$ regime. If $\xi$ is  large perturbation breaks down
and a more general treatment is required.

\section{Conclusions}\label{sec6}

In the present work we have examined the  low energy implications of F-theory $SU(5)\times U(1)' $ GUT models
 embedded in $SU(5)\times SU(5)'\supset SU(5)\times U(1)^4$. This gauge symmetry  emerges  naturally  from a single 
  point of $E_8$ enhancement, associated with the maximal geometric singularity  appearing in the elliptic
 fibration of the internal manifold. In order  to ensure realistic fermion mass textures and a tree-level top quark 
 Yukawa coupling,  we have imposed  a $Z_2$ monodromy group which acts on the geometric configuration of 7-branes 
 and identifies two out of the four abelian factors descending  from the $SU(5)'$ reduction. The $U(1)'$ symmetry of  the 
so derived effective field theory models, is a linear 
combination of the three remaining abelian symmetries descending from  $SU(5)'$. Imposing anomaly cancellation
conditions we have constructed all possible $U(1)'$ combinations and 
found as a generic property the appearance of non-universal $Z'$-couplings to 
the three families of quarks and leptons. 
Introducing fluxes consistent with 
the anomaly cancellation conditions, and  letting the 
 various neutral singlet-fields  acquire non-zero vevs,
we  obtained various effective models   distinguished from each other by their different low energy  spectra. 
We have  focused on viable classes of models derived in this framework. We have investigated the predictions 
on flavour changing currents and other processes mediated by the $Z'$ neutral gauge boson 
associated with the $U(1)'$ symmetry, which is supposed to break at some low energy scale.
Using the bounds on such processes coming from current investigation at LHC and other related experiments
we converted them to lower bounds on various parameters of the effective theory and in particular the $Z'$ mass. 
The present work provides a comprehensive classification of semi-local effective F-theory constructions 
reproducing the MSSM spectrum either with or without vector-like fields. On the phenomenological side,  
  the  focus  is  mainly in explorations of models with the   MSSM fields accompanied by several 
 neutral  singlets.  Fifty four  (54) MSSM models have been obtained 
and are classified  with respect to their predictions on the  $U(1)^{\prime}$ charges of the MSSM matter content.
In most of these cases,  $U(1)^{\prime}$ couples non-universally to the first two fermion families, 
and consequently  the $K_{0}-\overline{K_{0}}$ oscillation system forces  the strongest bound on the $Z^{\prime}$ mass.
As such, assuming reasonable values of the $U(1)^{\prime}$ gauge coupling $g^{\prime}$ we obtain $M_{Z^{\prime}}$ bounds at few hundreds TeV, well above the most recent LHC searches.
In other occasions  various flavour violation processes are predicted that can be tested on the ongoing or future experiments. 
The dominant process mediated by $Z'$ is the  lepton flavour violating  $\mu\rightarrow{eee}$ decay,
whilst its associated $\mu \to e \gamma$ rare reaction remains highly supperessed. 
Future  experiments designed to probe the lepton flavour violating process $\mu\rightarrow{eee}$   are expected to increase their sensitivity at about four orders of magnitude compared to the recent bounds. In this case the  models analysed in the present work are a spearhead for the interpretations
 of a positive experimental outcome. Even in
 the absence of any signal,  the foreseen bounds from $\mu\rightarrow{eee}$ searches will be  compatible with, if not dominant compared to the current
  bounds obtained in our models from neutral Kaon oscillation effects. 
On the other hand, we have seen that, models with $Z'$ coupled non-universally but only with MSSM spectrum, are not capable to interpret  the recently observed LHCb B-meson anomalies. All the same, our classification includes a class of models with vector-like families with non-trivial $Z'$-couplings which are capable to account for such effects.
These models display a universal 
nature of the $Z'$ couplings to the first two families with negligible
contributions to $K_{0}-\overline{K_{0}}$ oscillations. Their main feature is that the  $U(1)'$ charges of the vector-like fields differ
from those of the first two generations inducing this way   non-trivial
mixng effects. 
As an example, we briefly described such a model which includes a complete family of vector-like of fields where the observed LHCb B-meson anomalies can be explained through the mixing of the extra fermions with the  three generations of the SM. A detailed investigation of the whole class 
of these models will be presented in a future publication.

\noindent \\

\noindent \\

 \noindent{\Large{\textbf{Acknowledgements}} }
 
\noindent This research is co-financed by Greece and the European Union (European Social Fund-ESF) through the Operational Programme "Human Resources Development, Education and Lifelong Learning 2014-2020" in the context of the project "Grand Unified Theories from Superstring Theory: Theoretical Predictions and Modern Particle Physics Experiments" (MIS $\emph{5047638}$).

\newpage 
\begin{appendices}

\section{Anomaly Conditions: Analytic expressions}\label{appA}

 return Up to overall factors,  our computations give: $\mathcal{A}_{221}=\mathcal{A}_{331}=\mathcal{A}_{YY1}\equiv\mathcal{A}$, with 

\ba\label{A331}
\nonumber
\mathcal{A}&=&\left(30 \sqrt{3} c_1+15 \sqrt{6} c_2+9 \sqrt{10} c_3\right) m_1+\left(-60 \sqrt{3} c_1+15 \sqrt{6} c_2+9
   \sqrt{10} c_3\right) m_2+\left(9 \sqrt{10} c_3-45 \sqrt{6} c_2\right) m_3\\\nonumber
   &-& 36 \sqrt{10} c_3
   m_4 
   +\left(-20 \sqrt{3} c_1-10 \sqrt{6} c_2-6 \sqrt{10} c_3\right) M_1+\left(10 \sqrt{3} c_1-10
   \sqrt{6} c_2-6 \sqrt{10} c_3\right) M_2 \\
   &+& \left(-10 \sqrt{3} c_1+10 \sqrt{6} c_2-6 \sqrt{10} c_3\right)
   M_3+\left(-10 \sqrt{3} c_1-5 \sqrt{6} c_2+9 \sqrt{10} c_3\right) M_4\\\nn
   &+& \left(20 \sqrt{3} c_1+10 \sqrt{6}
   c_2-6 \sqrt{10} c_3\right) M_5 
   +\left(20 \sqrt{3} c_1-5 \sqrt{6} c_2+9 \sqrt{10} c_3\right)
   M_6+\left(15 \sqrt{6} c_2+9 \sqrt{10} c_3\right) M_7 \\\nn
   &+& 30 \sqrt{3} c_1 N_7+\left(10 \sqrt{3} c_1+20
   \sqrt{6} c_2\right) N_8+\left(10 \sqrt{3} c_1+5 \sqrt{6} c_2+15 \sqrt{10} c_3\right) N_9\; .\nn
   \ea

For the mixed $\mathcal{A}_{Y11}$ anomaly we have:
 
\be\label{AY11}
\begin{split}
 \mathcal{A}_{Y11}\;&=\;\frac{3}{2} \sqrt{\frac{3}{5}} c_1^2 N_7+\frac{1}{30} \left(\sqrt{15} c_1^2+4 \sqrt{30} c_2 c_1 +8
   \sqrt{15} c_2^2\right) N_{8}\\ 
  \;&+\;\frac{1}{60} \left(2 \sqrt{15} c_1^2+2 \sqrt{30} c_2 c_1+30 \sqrt{2} c_3
   c_1+\sqrt{15} c_2^2+15 \sqrt{15} c_3^2+30 c_2 c_3\right) N_9
   \end{split}
\ee

The $U(1)^{\prime}$-gravity anomaly yields the following expression:
  
    \ba\nn  \label{Agravity}
\mathcal{A}_{G}&=& \left(20 \sqrt{3} c_1+10 \sqrt{6} c_2+6 \sqrt{10} c_3\right) m_1+\left(-40 \sqrt{3} c_1+10 \sqrt{6} c_2+6
   \sqrt{10} c_3\right) m_2+\left(6 \sqrt{10} c_3-30 \sqrt{6} c_2\right) m_3\\\nn 
   &-&24 \sqrt{10} c_3
   m_4+\left(-20 \sqrt{3} c_1-10 \sqrt{6} c_2-6 \sqrt{10} c_3\right) M_1+\left(10 \sqrt{3} c_1-10
   \sqrt{6} c_2-6 \sqrt{10} c_3\right) M_2 \\\nn
   &+&\left(-10 \sqrt{3} c_1+10 \sqrt{6} c_2-6 \sqrt{10} c_3\right)
   M_3+\left(-10 \sqrt{3} c_1-5 \sqrt{6} c_2+9 \sqrt{10} c_3\right) M_4 \\\nn
   &+&\left(20 \sqrt{3} c_1+10 \sqrt{6}
   c_2-6 \sqrt{10} c_3\right) M_5 
   +\left(20 \sqrt{3} c_1-5 \sqrt{6} c_2+9 \sqrt{10} c_3\right)
   M_6+\left(15 \sqrt{6} c_2+9 \sqrt{10} c_3\right) M_7\\
   &+&24 \sqrt{3} c_1 N_7+\left(8 \sqrt{3} c_1+16
   \sqrt{6} c_2\right) N_8+\left(8 \sqrt{3} c_1+4 \sqrt{6} c_2+12 \sqrt{10} c_3\right) N_9 +\sum_{i\neq{j}}M_{ij}Q'_{ij}\; .
\ea

\noindent The pure cubic $U(1)^{\prime}$ anomaly is:
   
   \begin{small}
 \ba\nn \label{A111}
\mathcal{A}_{111}&=&\left(20 \sqrt{3} c_1^3+6 \left(5 \sqrt{6} c_2+3 \sqrt{10} c_3\right) c_1^2+6 \left(5 \sqrt{3} c_2^2+6
   \sqrt{5} c_3 c_2+3 \sqrt{3} c_3^2\right) c_1\right. \\ \nn
   &+&\left. 5 \sqrt{6} c_2^3+9 \sqrt{\frac{2}{5}} c_3^3+9 \sqrt{6} c_2
   c_3^2+9 \sqrt{10} c_2^2 c_3\right)m_{1} \\\nn
  &+&\left( -160 \sqrt{3} c_1^3+24 \left(5 \sqrt{6} c_2+3 \sqrt{10} c_3\right) c_1^2-12 \left(5 \sqrt{3} c_2^2+6 \sqrt{5}
   c_3 c_2+3 \sqrt{3} c_3^2\right) c_{1}\right. \\ \nn
  &+&\left. 5 \sqrt{6} c_2^3+9 \sqrt{\frac{2}{5}} c_3^3+9 \sqrt{6} c_2 c_3^2+9\sqrt{10} c_2^2 c_3\right)m_2 \\ \nn
  &-&9\left(15 \sqrt{6} c_2^3-9 \sqrt{10} c_3 c_2^2+3 \sqrt{6} c_3^2 c_2- \sqrt{\frac{2}{5}} c_3^3  \right)m_3 -576\sqrt{\frac{2}{5}}c_{3}^{2} m_4\\ \nn
  &-& \left(80 \sqrt{3} c_1^3+24 \left(5 \sqrt{6} c_2+3 \sqrt{10} c_3\right) c_1^2+24 \left(5
   \sqrt{3} c_2^2+6 \sqrt{5} c_3 c_2+3 \sqrt{3} c_3^2\right) c_1 \right. \\\nn 
   &+& \left. 20 \sqrt{6} c_2^3+9 \sqrt{10} c_3^3+36
   \sqrt{6} c_2 c_3^2+36 \sqrt{10} c_2^2 c_3\right)M_{0}\\ \nn
   &+&\left( 10 \sqrt{3} c_1^3-6 \left(5 \sqrt{6} c_2+3 \sqrt{10} c_3\right) c_1^2+12 \left(5 \sqrt{3} c_2^2+6
   \sqrt{5} c_3 c_2+3 \sqrt{3} c_3^2\right) c_1 \right. \\\nn
   &-&\left. 20 \sqrt{6} c_2^3-36 \sqrt{10} c_3
   c_2^2-36 \sqrt{6} c_3^2 c_2-36 \sqrt{\frac{2}{5}} c_3^3\right)M_1\\ \nn
   &-&\left( 10 \sqrt{3} c_1^3+6 \left(5 \sqrt{6} c_2-3 \sqrt{10} c_3\right) c_1^2-12 \left(5 \sqrt{3} c_2^2-6
   \sqrt{5} c_3 c_2+3 \sqrt{3} c_3^2\right) c_1\right.\\\nn
   &+&\left.20 \sqrt{6} c_2^3-36 \sqrt{\frac{2}{5}} c_3^3+36 \sqrt{6}
   c_2 c_3^2-36 \sqrt{10} c_2^2 c_3\right)M_2\\ \nn
   &-&\left( 10 \sqrt{3} c_1^3-3 \left(5 \sqrt{6} c_2-9 \sqrt{10} c_3\right) c_1^2-3 \left(5 \sqrt{3} c_2^2-18
   \sqrt{5} c_3 c_2+27 \sqrt{3} c_3^2\right) c_1\right. \\\nn
   &-&\left. 5 \sqrt{\frac{3}{2}} c_2^3+\frac{243
   c_3^3}{\sqrt{10}}-81 \sqrt{\frac{3}{2}} c_2 c_3^2+27 \sqrt{\frac{5}{2}} c_2^2 c_3\right)M_3\\\nn
   &+&\left( 80 \sqrt{3} c_1^3+20 \sqrt{6} c_2^3-36 \sqrt{\frac{2}{5}} c_3^3+36 \sqrt{6} c_2 c_3^2-36 \sqrt{10} c_2^2 c_3\right.\\\nn
   &+&\left. 24 \left(5 \sqrt{6} c_2-3 \sqrt{10} c_3\right) c_1^2+24 \left(5 \sqrt{3} c_2^2-6 \sqrt{5} c_3 c_2+3 \sqrt{3} c_3^2\right) c_1\right)M_4\\\nn
   &+& \left(80 \sqrt{3} c_1^3-12 \left(5 \sqrt{6} c_2-9 \sqrt{10} c_3\right) c_1^2+6 \left(5 \sqrt{3} c_2^2-18
   \sqrt{5} c_3 c_2+27 \sqrt{3} c_3^2\right) c_1\right.\\\nn 
   &-&\left. 5 \sqrt{\frac{3}{2}} c_2^3+\frac{243
   c_3^3}{\sqrt{10}}-81 \sqrt{\frac{3}{2}} c_2 c_3^2+27 \sqrt{\frac{5}{2}} c_2^2 c_3\right)M_5\\
   &+&\frac{27}{10} \left(25 \sqrt{6} c_2^3+45 \sqrt{10} c_3 c_2^2+45 \sqrt{6} c_3^2 c_2+9 \sqrt{10}
   c_3^3\right)M_6  +\sum_{i\neq{j}}M_{ij}Q^{\prime \; 3}_{ij}
   \ea
   \end{small}

\noindent The last terms in \eqref{Agravity} and \eqref{A111} represents the contribution from the singlets.

\section{List of models}\label{AppB}
In this Appendix all the flux solutions subject to MSSM spectrum criteria, the corresponding $U(1)^{\prime}$-charges and details about the singlet spectrum are presented. For each $c_{i}$-solution presented, a similar solution subject to $c_{i}\rightarrow{-c_{i}}$ is also predicted from the solution of the anomaly cancellation conditions. Hence, models with charges subject to $Q^{\prime}\rightarrow{-Q^{\prime}}$ are also exist.

As mentioned on the main text, there are fifty-four solutions that fall into four classes of models: \emph{Class} A, B, C and D.

\subsection*{Class A}

This class consists of six models. The flux data solutions along with the resulting $c_{i}$-coefficients have been presented in Table \ref{tab:classA_fluxes} of the main text. The corresponding models defined by these solutions  along with their $U(1)^{\prime}$ charges are given in Table \ref{tab:classA_models}. Here we present only the singlet spectrum for this class of models.

As have been already discussed, in this particular class of models the singlets come in pairs, meaning that $M_{ij}=M_{ji}$. Hence, a minimal singlet spectrum scenario implies that $M_{ij}=M_{ji}=1$. The singlet charges $Q_{ij}^{\prime}$ for each model are given in Table \ref{tab:classA_singlets}, below.

\begin{table}[H]
\centering\resizebox{0.53\textwidth}{!}{%
\begin{tabular}{l|cccccc}\hline\hline
\textbf{Class A} &   \multicolumn{6}{c}{Charges}        \\\hline
Models  & $Q_{13}^{\prime}$ & $Q_{14}^{\prime}$ & $Q_{15}^{\prime}$ & $Q_{34}^{\prime}$ & $Q_{35}^{\prime}$ & $Q_{45}^{\prime}$  \\\hline\hline
\textbf{A1}, \textbf{A6}  & 0  &  $\frac{1}{2}$  &  $-\frac{1}{2}$ & $\frac{1}{2}$ &  $-\frac{1}{2}$ &  $-1$ \\
\textbf{A2}, \textbf{A5}  & $\frac{1}{2}$  & 0  &  $-\frac{1}{2}$ & $-\frac{1}{2}$ & $-1$ &  $-\frac{1}{2}$ \\
\textbf{A3}, \textbf{A4}  & $-\frac{1}{2}$  & $\frac{1}{2}$ & $0$ & 1 & $\frac{1}{2}$ & $-\frac{1}{2}$  \\
\hline\hline
\end{tabular}}
\caption{ \small{Singlets charges of Class A models. }}
\label{tab:classA_singlets}
\end{table}

\subsection*{Class B}

This Class of models consists of twenty-four solutions. All the relevant data characterized the models organized in three tables. In particular, Table \ref{tab:classB_fluxes} contains the flux data of the models along with the corresponding $c_{i}$-solutions, as those have been extracted from the solution of the anomaly cancellation conditions. In Table \ref{tab:classB_charges}, the $U(1)^{\prime}$ charges of the matter curves are given. Finally, details about the singlet spectrum presented in Table \ref{tab:classB_singlets}.

\begin{table}[H]
\centering\resizebox{\textwidth}{!}{%
\begin{tabular}{l|cccccccccccccc|ccc}\hline\hline
\textbf{Class B} & \multicolumn{14}{c|}{Flux data}                                   & \multicolumn{3}{c}{$c_{i}$ coefficients}        \\\hline
Model  & $m_1$ & $m_2$ & $m_3$ & $m_4$ & $M_1$                                                                             & $M_2$ & $M_3$ & $M_4$ & $M_5$ & $M_6$ & $M_7$ & $N_7$ & $N_8$ & $N_9$                                                                                                                 & $c_1$ & $c_2$ & $c_3$  \\\hline\hline
\textbf{B1} & 1 & 0 & 1 & 1 & 0 & -1 & 0 & 0 & 0 & -1 & -1 & 0 & 0 & 1 & -$\frac{\sqrt{5}}{3}$ & $-\frac{\sqrt{\frac{5}{2}}}{3}$ & $\frac{1}{\sqrt{6}}$ \\
\textbf{B2} & 1 & 0 & 1 & 1 & 0 & 0 & -1 & 0 & 0 & -1 & -1 & 0 & 0 & 1 & $-\frac{\sqrt{5}}{3}$ & $-\frac{\sqrt{\frac{5}{2}}}{3}$ & $\frac{1}{\sqrt{6}}$ \\
\textbf{B3} & 1 & 0 & 1 & 1 & 0 & 0 & 0 & -1 & -1 & 0 & -1 & 0 & 1 & 0 & $\frac{\sqrt{5}}{3}$ & $-\frac{\sqrt{\frac{5}{2}}}{6}$ & $\frac{\sqrt{\frac{3}{2}}}{2}$ \\
\textbf{B4} & 1 & 0 & 1 & 1 & 0 & 0 & 0 & 0 & -2 & 0 & -1 & 0 & 1 & 0 & $\frac{\sqrt{5}}{3}$ & $-\frac{\sqrt{\frac{5}{2}}}{6}$ & $\frac{\sqrt{\frac{3}{2}}}{2}$ \\
\textbf{B5} & 1 & 0 & 1 & 1 & 0 & 0 & 0 & 0 & -1 & 0 & -2 & 0 & 1 & 0 & $\frac{\sqrt{5}}{3}$ & $-\frac{\sqrt{\frac{5}{2}}}{6}$ & $\frac{\sqrt{\frac{3}{2}}}{2}$ \\
\textbf{B6} & 1 & 0 & 1 & 1 & 0 & 0 & 0 & 0 & 0 & -2 & -1 & 0 & 0 & 1 & $-\frac{\sqrt{5}}{3}$ & $-\frac{\sqrt{\frac{5}{2}}}{3}$ & $\frac{1}{\sqrt{6}}$ \\
\textbf{B7} & 1 & 0 & 1 & 1 & 0 & -1 & 0 & 0 & -1 & 0 & -1 & 0 & 1 & 0 & $\frac{\sqrt{5}}{3}$ & $-\frac{\sqrt{\frac{5}{2}}}{6}$ & $\frac{\sqrt{\frac{3}{2}}}{2}$ \\
\textbf{B8} & 1 & 0 & 1 & 1 & 0 & 0 & 0 & 0 & 0 & -1 & -2 & 0 & 0 & 1 & $-\frac{\sqrt{5}}{3}$ & $-\frac{\sqrt{\frac{5}{2}}}{3}$ & $\frac{1}{\sqrt{6}}$ \\
\textbf{B9} & 1 & 1 & 0 & 1 & 0 & -1 & 0 & 0 & 0 & -1 & -1 & 0 & 0 & 1 & $-\frac{\sqrt{5}}{3}$ & $-\frac{\sqrt{\frac{5}{2}}}{3}$ & $\frac{1}{\sqrt{6}}$ \\
\textbf{B10} & 1 & 1 & 0 & 1 & 0 & 0 & -1 & 0 & -1 & -1 & 0 & 1 & 0 & 0 & 0 & $-\frac{\sqrt{\frac{5}{2}}}{2}$ & $-\frac{\sqrt{\frac{3}{2}}}{2}$ \\
\textbf{B11} & 1 & 1 & 0 & 1 & 0 & 0 & -1 & 0 & 0 & -1 & -1 & 0 & 0 & 1 & $-\frac{\sqrt{5}}{3}$ & $-\frac{\sqrt{\frac{5}{2}}}{3}$ & $\frac{1}{\sqrt{6}}$ \\
\textbf{B12} & 1 & 1 & 0 & 1 & 0 & 0 & 0 & -1 & -1 & -1 & 0 & 1 & 0 & 0 & 0 & $-\frac{\sqrt{\frac{5}{2}}}{2}$ & $-\frac{\sqrt{\frac{3}{2}}}{2}$\\
\textbf{B13} & 1 & 1 & 0 & 1 & 0 & 0 & 0 & 0 & -2 & -1 & 0 & 1 & 0 & 0 & 0 & $-\frac{\sqrt{\frac{5}{2}}}{2}$ & $-\frac{\sqrt{\frac{3}{2}}}{2}$ \\
\textbf{B14} & 1 & 1 & 0 & 1 & 0 & 0 & 0 & 0 & -1 & -2 & 0 & 1 & 0 & 0 & 0 & $-\frac{\sqrt{\frac{5}{2}}}{2}$ & $-\frac{\sqrt{\frac{3}{2}}}{2}$ \\
\textbf{B15} & 1 & 1 & 0 & 1 & 0 & 0 & 0 & 0 & 0 & -2 & -1 & 0 & 0 & 1 & $-\frac{\sqrt{5}}{3}$ & $-\frac{\sqrt{\frac{5}{2}}}{3}$ & $\frac{1}{\sqrt{6}}$ \\
\textbf{B16} & 1 & 1 & 0 & 1 & 0 & 0 & 0 & 0 & 0 & -1 & -2 & 0 & 0 & 1 & $-\frac{\sqrt{5}}{3}$ & $-\frac{\sqrt{\frac{5}{2}}}{3}$ & $\frac{1}{\sqrt{6}}$ \\
\textbf{B17} & 1 & 1 & 1 & 0 & 0 & -1 & 0 & 0 & -1 & 0 & -1 & 0 & 1 & 0 & $\frac{\sqrt{5}}{3}$ & $-\frac{\sqrt{\frac{5}{2}}}{6}$ & $\frac{\sqrt{\frac{3}{2}}}{2}$ \\
\textbf{B18} & 1 & 1 & 1 & 0 & 0 & 0 & -1 & 0 & -1 & -1 & 0 & 1 & 0 & 0 & 0 & $-\frac{\sqrt{\frac{5}{2}}}{2}$ & $-\frac{\sqrt{\frac{3}{2}}}{2}$ \\
\textbf{B19} & 1 & 1 & 1 & 0 & 0 & 0 & 0 & -1 & -1 & -1 & 0 & 1 & 0 & 0 & 0 & $-\frac{\sqrt{\frac{5}{2}}}{2}$ & $-\frac{\sqrt{\frac{3}{2}}}{2}$ \\
\textbf{B20} & 1 & 1 & 1 & 0 & 0 & 0 & 0 & -1 & -1 & 0 & -1 & 0 & 1 & 0 & $\frac{\sqrt{5}}{3}$ & $-\frac{\sqrt{\frac{5}{2}}}{6}$ & $\frac{\sqrt{\frac{3}{2}}}{2}$ \\
\textbf{B21} & 1 & 1 & 1 & 0 & 0 & 0 & 0 & 0 & -2 & -1 & 0 & 1 & 0 & 0 & 0 & $-\frac{\sqrt{\frac{5}{2}}}{2}$ & $-\frac{\sqrt{\frac{3}{2}}}{2}$ \\
\textbf{B22} & 1 & 1 & 1 & 0 & 0 & 0 & 0 & 0 & -2 & 0 & -1 & 0 & 1 & 0 & $\frac{\sqrt{5}}{3}$ & $-\frac{\sqrt{\frac{5}{2}}}{6}$ & $\frac{\sqrt{\frac{3}{2}}}{2}$ \\
\textbf{B23} & 1 & 1 & 1 & 0 & 0 & 0 & 0 & 0 & -1 & -2 & 0 & 1 & 0 & 0 & 0 & $-\frac{\sqrt{\frac{5}{2}}}{2}$ & $-\frac{\sqrt{\frac{3}{2}}}{2}$ \\
\textbf{B24} & 1 & 1 & 1 & 0 & 0 & 0 & 0 & 0 & -1 & 0 & -2 & 0 & 1 & 0 & $\frac{\sqrt{5}}{3}$ & $-\frac{\sqrt{\frac{5}{2}}}{6}$ & $\frac{\sqrt{\frac{3}{2}}}{2}$ \\\hline\hline
	\end{tabular}}
	\caption{ \small{Class B models, flux data and the corresponding $c_{i}$-solutions.}}
	\label{tab:classB_fluxes}
	\end{table}

\begin{table}[H]
\centering\resizebox{\textwidth}{!}{%
\begin{tabular}{l|cccc|ccccccc}\hline\hline
\textbf{Class B} & \multicolumn{11}{c}{Charges$\times{\sqrt{15}}$}                                          \\\hline
Models  &  $Q^{\prime}_{10_{1}}$ &  $Q^{\prime}_{10_{2}}$ &  $Q^{\prime}_{10_{3}}$ &  $Q^{\prime}_{10_{4}}$ &  $Q^{\prime}_{5_{1}}$                                                                             &  $Q^{\prime}_{5_{2}}$ &  $Q^{\prime}_{5_{3}}$&  $Q^{\prime}_{5_{4}}$ &  $Q^{\prime}_{5_{5}}$ &  $Q^{\prime}_{5_{6}}$ &  $Q^{\prime}_{5_{7}}$                                                                                               \\\hline\hline
\textbf{B1}, \textbf{B2}, \textbf{B6}, \textbf{B8}, \textbf{B9}, \textbf{B11}, \textbf{B15}, \textbf{B16} & -1 & 3/2 & 3/2 & -1 & 2 & -1/2 & -1/2 & 2 & -3 & -1/2 & -1/2 \\
\textbf{B3}, \textbf{B4}, \textbf{B5}, \textbf{B7},  \textbf{B17}, \textbf{B20}, \textbf{B22}, \textbf{B24} & 1 & -3/2 & 1 & -3/2 & -2 & 1/2 & -2 & 1/2 & 1/2 & 3 & 1/2 \\
\textbf{B10}, \textbf{B12}, \textbf{B13}, \textbf{B14}, \textbf{B18}, \textbf{B19}, \textbf{B21}, \textbf{B23} & -1 & -1 & 3/2 & 3/2 & 2 & 2 & -1/2 & -1/2 & -1/2 & -1/2 & -3 \\
 \hline\hline
	\end{tabular}}
	\caption{ \small{$U(1)^{\prime}$ charges of Class B models. }}
	\label{tab:classB_charges}
	\end{table}

	\begin{table}[H]
\centering\resizebox{\textwidth}{!}{%
\begin{tabular}{l|cccccccccccc|cccccc}\hline\hline
\textbf{Class B} & \multicolumn{12}{c|}{Multiplicities}                                   & \multicolumn{6}{c}{Charges$\times{\sqrt{15}}$}        \\\hline
Models  & $M_{13}$ & $M_{14}$ & $M_{15}$ & $M_{34}$ & $M_{35}$ & $M_{45}$ & $M_{31}$ & $M_{41}$ & $M_{51}$ & $M_{43}$ & $M_{53}$ & $M_{54}$ & $Q_{13}^{\prime}$ & $Q_{14}^{\prime}$ & $Q_{15}^{\prime}$ & $Q_{34}^{\prime}$ & $Q_{35}^{\prime}$ & $Q_{45}^{\prime}$  \\\hline\hline
\textbf{B1}, \textbf{B2}, \textbf{B6},     &   &  &  &  &  &  &  &  &  &  &  &  & &  &  & & &  \\
\textbf{B8}, \textbf{B9}, \textbf{B11},      &  1 & 2 & 2 & 1 & 1 & 1 & 1 & 1 & 1 & 1 & 1 & 1 &$-\frac{5}{2}$ & $-\frac{5}{2}$ & 0 & 0 & $\frac{5}{2}$ & $\frac{5}{2}$   \\
\textbf{B15}, \textbf{B16}  &   &  &  &  &  &  &  &  &  &  &  &  & &  &  & & &  \\\hline
\textbf{B3}, \textbf{B4}, \textbf{B5},     &   &  &  &  &  &  &  &  &  &  &  &  & &  &  & & &  \\
\textbf{B7}, \textbf{B17}, \textbf{B20},      &  1 & 2 & 2 & 1 & 1 & 1 & 1 & 1 & 1 & 1 & 1 & 1 &$\frac{5}{2}$ & 0 & $\frac{5}{2}$  &  $-\frac{5}{2}$ & 0 & $\frac{5}{2}$   \\
\textbf{B22}, \textbf{B24}  &   &  &  &  &  &  &  &  &  &  &  &  & &  &  & & &  \\\hline
\textbf{B10}, \textbf{B12}, \textbf{B13},     &   &  &  &  &  &  &  &  &  &  &  &  & &  &  & & &  \\
\textbf{B14}, \textbf{B18}, \textbf{B19},      &  1 & 2 & 2 & 1 & 1 & 1 & 1 & 1 & 1 & 1 & 1 & 1 &0 & $-\frac{5}{2}$ & $-\frac{5}{2}$  &  $-\frac{5}{2}$ & $-\frac{5}{2}$ & 0  \\
\textbf{B21}, \textbf{B23}  &   &  &  &  &  &  &  &  &  &  &  &  & &  &  & & & 
 \\\hline\hline
\end{tabular}}
\caption{ \small{Singlets spectrum of Class B models. }}
\label{tab:classB_singlets}
\end{table}

\subsection*{Class C}

Twelve models define this class. Gauge anomaly cancellation solutions are given in Table \ref{tab:classC_fluxes} while the corresponding matter curve $U(1)^{\prime}$ charges are listed in Table \ref{tab:classC_charges}. The properties of the singlet spectrum are described in Table \ref{tab:classC_singlets}. 

\begin{table}[H]
\centering\resizebox{\textwidth}{!}{%
\begin{tabular}{l|cccccccccccccc|ccc}\hline\hline
\textbf{Class C} & \multicolumn{14}{c|}{Flux data}                                   & \multicolumn{3}{c}{$c_{i}$ coefficients}        \\\hline
Model  & $m_1$ & $m_2$ & $m_3$ & $m_4$ & $M_1$                                                                             & $M_2$ & $M_3$ & $M_4$ & $M_5$ & $M_6$ & $M_7$ & $N_7$ & $N_8$ & $N_9$                                                                                                                 & $c_1$ & $c_2$ & $c_3$  \\\hline\hline
\textbf{C1} & 1 & 0 & 0 & 2 & 0 & -1 & 0 & 0 & 0 & -1 & -1 & 0 & 0 & 1 & $-\frac{\sqrt{5}}{3}$ & $\frac{5 \sqrt{\frac{5}{2}}}{12}$ & $\frac{1}{4 \sqrt{6}}$ \\
\textbf{C2} & 1 & 0 & 0 & 2 & 0 & 0 & 0 & 0 & 0 & -2 & -1 & 0 & 0 & 1 & $-\frac{\sqrt{5}}{3}$ & $\frac{5 \sqrt{\frac{5}{2}}}{12}$ & $\frac{1}{4 \sqrt{6}}$ \\
\textbf{C3} & 1 & 0 & 0 & 2 & 0 & 0 & 0 & 0 & 0 & -1 & -2 & 0 & 0 & 1 & $-\frac{\sqrt{5}}{6}$ & $\frac{7 \sqrt{\frac{5}{2}}}{12}$ & $-\frac{1}{4 \sqrt{6}}$ \\
\textbf{C4} & 1 & 0 & 2 & 0 & 0 & -1 & 0 & 0 & -1 & 0 & -1 & 0 & 1 & 0 & $\frac{\sqrt{5}}{3}$ & $-\frac{\sqrt{\frac{5}{2}}}{6}$ & $-\frac{\sqrt{\frac{3}{2}}}{2}$ \\
\textbf{C5} & 1 & 0 & 2 & 0 & 0 & 0 & 0 & -1 & -1 & 0 & -1 & 0 & 1 & 0 & $\frac{\sqrt{5}}{6}$ & $-\frac{\sqrt{\frac{5}{2}}}{12}$ & $-\frac{3 \sqrt{\frac{3}{2}}}{4}$ \\
\textbf{C6} & 1 & 0 & 2 & 0 & 0 & 0 & 0 & 0 & -2 & 0 & -1 & 0 & 1 & 0 & $\frac{\sqrt{5}}{3}$ & $-\frac{\sqrt{\frac{5}{2}}}{6}$ & $-\frac{\sqrt{\frac{3}{2}}}{2}$\\
\textbf{C7} & 1 & 0 & 2 & 0 & 0 & 0 & 0 & 0 & -1 & 0 & -2 & 0 & 1 & 0 & $\frac{\sqrt{5}}{6}$ & $-\frac{\sqrt{\frac{5}{2}}}{12}$ & $-\frac{3 \sqrt{\frac{3}{2}}}{4}$ \\
\textbf{C8} & 1 & 0 & 0 & 2 & 0 & 0 & -1 & 0 & 0 & -1 & -1 & 0 & 0 & 1 & $-\frac{\sqrt{5}}{6}$ & $\frac{7 \sqrt{\frac{5}{2}}}{12}$ & $-\frac{1}{4 \sqrt{6}}$ \\
\textbf{C9} & 1 & 2 & 0 & 0 & 0 & 0 & -1 & 0 & -1 & -1 & 0 & 1 & 0 & 0 & 0 & $\frac{\sqrt{\frac{5}{2}}}{2}$ & $-\frac{\sqrt{\frac{3}{2}}}{2}$ \\
\textbf{C10} & 1 & 2 & 0 & 0 & 0 & 0 & 0 & -1 & -1 & -1 & 0 & 1 & 0 & 0 & 0 & $\frac{\sqrt{\frac{5}{2}}}{4}$ & $-\frac{3 \sqrt{\frac{3}{2}}}{4}$ \\
\textbf{C11} & 1 & 2 & 0 & 0 & 0 & 0 & 0 & 0 & -2 & -1 & 0 & 1 & 0 & 0 & 0 & $\frac{\sqrt{\frac{5}{2}}}{2}$ & $-\frac{\sqrt{\frac{3}{2}}}{2}$ \\
\textbf{C12} & 1 & 2 & 0 & 0 & 0 & 0 & 0 & 0 & -1 & -2 & 0 & 1 & 0 & 0 & 0 & $\frac{\sqrt{\frac{5}{2}}}{4}$ & $-\frac{3 \sqrt{\frac{3}{2}}}{4}$ 
 \\\hline\hline
	\end{tabular}}
	\caption{ \small{Class C models, flux data along with the corresponding $c_{i}$-coefficients. }}
	\label{tab:classC_fluxes}
	\end{table}

\begin{table}[H]
\centering\resizebox{14cm}{!}{%
\begin{tabular}{l|cccc|ccccccc}\hline\hline
\textbf{Class C} & \multicolumn{11}{c}{Charges$\times{\sqrt{15}}$}                                          \\\hline
Models  &  $Q^{\prime}_{10_{1}}$ &  $Q^{\prime}_{10_{2}}$ &  $Q^{\prime}_{10_{3}}$ &  $Q^{\prime}_{10_{4}}$ &  $Q^{\prime}_{5_{1}}$                                                                             &  $Q^{\prime}_{5_{2}}$ &  $Q^{\prime}_{5_{3}}$&  $Q^{\prime}_{5_{4}}$ &  $Q^{\prime}_{5_{5}}$ &  $Q^{\prime}_{5_{6}}$ &  $Q^{\prime}_{5_{7}}$                                                                                               \\\hline\hline
\textbf{C1}, \textbf{C2} & -1/4 & 9/4 & -3/2 & -1/4 & 1/2 & -2 & 7/4 & 1/2 & -3/4 & -2 & 7/4 \\
\textbf{C3}, \textbf{C8} & 1/4 & 3/2 & -9/4 & 1/4 & -1/2 & -7/4 & 2 & -1/2 & 3/4 & -7/4 & 2 \\
\textbf{C4}, \textbf{C6} & 1/4 & -9/4 & 1/4 & 3/2 & -1/2 & 2 & -1/2 & -7/4 & 2 & 3/4 & -7/4 \\
\textbf{C5}, \textbf{C7} & -1/4 & -3/2 & -1/4 & 9/4 & 1/2 & 7/4 & 1/2 & -2 & 7/4 & -3/4 & -2 \\
\textbf{C9}, \textbf{C11} & 1/4 & 1/4 & -9/4 & 3/2 & -1/2 & -1/2 & 2 & -7/4 & 2 & -7/4 & 3/4 \\
\textbf{C10}, \textbf{C12} & -1/4 & -1/4 & -3/2 & 9/4 & 1/2 & 1/2 & 7/4 & -2 & 7/4 & -2 & -3/4 \\
 \hline\hline
	\end{tabular}}
	\caption{ \small{$U(1)^{\prime}$ charges of Class C models. The charges are multiplied with $\sqrt{15}$. }}
	\label{tab:classC_charges}
	\end{table}

\begin{table}[H]
\centering\resizebox{\textwidth}{!}{%
\begin{tabular}{l|cccccccccccc|cccccc}\hline\hline
\textbf{Class C} & \multicolumn{12}{c|}{Multiplicities}                                   & \multicolumn{6}{c}{Charges$\times{\sqrt{15}}$}        \\\hline
Models  & $M_{13}$ & $M_{14}$ & $M_{15}$ & $M_{34}$ & $M_{35}$ & $M_{45}$ & $M_{31}$ & $M_{41}$ & $M_{51}$ & $M_{43}$ & $M_{53}$ & $M_{54}$ & $Q_{13}^{\prime}$ & $Q_{14}^{\prime}$ & $Q_{15}^{\prime}$ & $Q_{34}^{\prime}$ & $Q_{35}^{\prime}$ & $Q_{45}^{\prime}$  \\\hline\hline
\textbf{C1}, \textbf{C2}      &  1 & 1 & 1 & 1 & 1 & 1 & 1 & 1 & 1 & 2 & 1 & 1 &$-\frac{5}{2}$ & $\frac{5}{4}$ & 0 & $\frac{15}{4}$ & $\frac{5}{2}$ & $-\frac{5}{4}$   \\
\textbf{C3}, \textbf{C8}  & 1 & 1 & 1 & 2 & 1 & 1 & 1 & 1 & 1 & 1 & 1 & 1  &$-\frac{5}{4}$ & $\frac{5}{2}$ & 0 & $\frac{15}{4}$ & $\frac{5}{4}$ & $-\frac{5}{2}$   \\
\textbf{C4}, \textbf{C6} & 1 & 1 & 1 & 1 & 1 & 1 & 1 & 1 & 1 & 1 & 2 & 1 &$\frac{5}{2}$ & 0 &  $-\frac{5}{4}$ & $-\frac{5}{2}$ & $-\frac{15}{4}$ & $-\frac{5}{4}$   \\
\textbf{C5}, \textbf{C7} & 1 & 1 & 1 & 1 & 2 & 1 & 1 & 1 & 1 & 1 & 1 & 1 &$\frac{5}{4}$ & 0 &  $-\frac{5}{2}$ & $-\frac{5}{4}$ & $-\frac{15}{4}$ & $-\frac{5}{2}$   \\
\textbf{C9}, \textbf{C11} & 1 & 1 & 1 & 1 & 1 & 1 & 1 & 1 & 1 & 1 & 1 & 2 &0 & $\frac{5}{2}$ &  $-\frac{5}{4}$ & $\frac{5}{2}$ & $-\frac{5}{4}$ & $-\frac{15}{4}$   \\
\textbf{C10}, \textbf{C12} & 1 & 1 & 1 & 1 & 1 & 2 & 1 & 1 & 1 & 1 & 1 & 1 & 0 & $\frac{5}{4}$ &  $-\frac{5}{2}$ & $\frac{5}{4}$ & $-\frac{5}{2}$ & $-\frac{15}{4}$ \\\hline\hline
\end{tabular}}
\caption{ \small{Singlets spectrum of Class C models. }}
\label{tab:classC_singlets}
\end{table}

\subsection*{Class D}

This class contains twelve models. Flux data along with the corresponding solution for the $c_{i}$-coefficients are given in Table \ref{tab:classD_fluxes}. The $U(1)^{\prime}$ charges are listed in Table \ref{tab:classD_charges} while the properties (multiplicities and $Q_{ij}^{\prime}$ charges) of the singlet spectrum are described in Table \ref{tab:classD_singlets}.

\begin{table}[H]
\centering\resizebox{\textwidth}{!}{%
\begin{tabular}{l|cccccccccccccc|ccc}\hline\hline
\textbf{Class D} & \multicolumn{14}{c|}{Flux data}                                   & \multicolumn{3}{c}{$c_{i}$ coefficients}        \\\hline
Model  & $m_1$ & $m_2$ & $m_3$ & $m_4$ & $M_1$                                                                             & $M_2$ & $M_3$ & $M_4$ & $M_5$ & $M_6$ & $M_7$ & $N_7$ & $N_8$ & $N_9$                                                                                                                 & $c_1$ & $c_2$ & $c_3$  \\\hline\hline
\textbf{D1} & 1 & 0 & 1 & 1 & 0 & 0 & -1 & 0 & -1 & 0 & -1 & 0 & 1 & 0 & $\frac{\sqrt{\frac{5}{6}}}{2}$ & $-\frac{\sqrt{\frac{5}{3}}}{8}$ & $\frac{7}{8}$ \\
\textbf{D2} & 1 & 0 & 1 & 1 & 0 & 0 & 0 & -1 & 0 & -1 & -1 & 0 & 0 & 1 & $\frac{\sqrt{\frac{5}{6}}}{2}$ & $\frac{5 \sqrt{\frac{5}{3}}}{8}$ & $-\frac{3}{8}$ \\
\textbf{D3} & 1 & 0 & 1 & 1 & 0 & 0 & 0 & 0 & -1 & -1 & -1 & 0 & 0 & 1 & $-\sqrt{\frac{5}{6}}$ & $-\frac{\sqrt{\frac{5}{3}}}{8}$ & $\frac{3}{8}$ \\
\textbf{D4} & 1 & 0 & 1 & 1 & 0 & 0 & 0 & 0 & -1 & -1 & -1 & 0 & 1 & 0 & $\sqrt{\frac{5}{6}}$ & $-\frac{\sqrt{\frac{5}{3}}}{4}$ & $\frac{1}{4}$ \\
\textbf{D5} & 1 & 1 & 0 & 1 & 0 & -1 & 0 & 0 & -1 & -1 & 0 & 1 & 0 & 0 & 0 & $-\frac{\sqrt{15}}{8}$ & $-\frac{7}{8}$ \\
\textbf{D6} & 1 & 1 & 0 & 1 & 0 & 0 & 0 & -1 & 0 & -1 & -1 & 0 & 0 & 1 & $-\sqrt{\frac{5}{6}}$ & $-\frac{\sqrt{\frac{5}{3}}}{8}$ & $\frac{3}{8}$ \\
\textbf{D7} & 1 & 1 & 0 & 1 & 0 & 0 & 0 & 0 & -1 & -1 & -1 & 1 & 0 & 0 & 0 & $-\frac{\sqrt{15}}{4}$ & $-\frac{1}{4}$\\
\textbf{D8} & 1 & 1 & 1 & 0 & 0 & -1 & 0 & 0 & -1 & -1 & 0 & 1 & 0 & 0 & 0 & $-\frac{\sqrt{15}}{4}$ & $-\frac{1}{4}$ \\
\textbf{D9} & 1 & 1 & 0 & 1 & 0 & 0 & 0 & 0 & -1 & -1 & -1 & 0 & 0 & 1 & $\frac{\sqrt{\frac{5}{6}}}{2}$ & $\frac{5 \sqrt{\frac{5}{3}}}{8}$ & $-\frac{3}{8}$\\
\textbf{D10} & 1 & 1 & 1 & 0 & 0 & 0 & -1 & 0 & -1 & 0 & -1 & 0 & 1 & 0 & $\sqrt{\frac{5}{6}}$ & $-\frac{\sqrt{\frac{5}{3}}}{4}$ & $\frac{1}{4}$\\
\textbf{D11} & 1 & 1 & 1 & 0 & 0 & 0 & 0 & 0 & -1 & -1 & -1 & 0 & 1 & 0 & $\frac{\sqrt{\frac{5}{6}}}{2}$ & $-\frac{\sqrt{\frac{5}{3}}}{8}$ & $\frac{7}{8}$ \\
\textbf{D12} & 1 & 1 & 1 & 0 & 0 & 0 & 0 & 0 & -1 & -1 & -1 & 1 & 0 & 0 & 0 & $-\frac{\sqrt{15}}{8}$ & $-\frac{7}{8}$
 \\\hline\hline
	\end{tabular}}
	\caption{ \small{Class D models flux data. }}
	\label{tab:classD_fluxes}
	\end{table} 

\begin{table}[H]
\centering\small
\begin{tabular}{l|cccc|ccccccc}\hline\hline
\textbf{Class D} & \multicolumn{11}{c}{Charges$\times{\sqrt{10}}$}                                          \\\hline
Models  &  $Q^{\prime}_{10_{1}}$ &  $Q^{\prime}_{10_{2}}$ &  $Q^{\prime}_{10_{3}}$ &  $Q^{\prime}_{10_{4}}$ &  $Q^{\prime}_{5_{1}}$                                                                             &  $Q^{\prime}_{5_{2}}$ &  $Q^{\prime}_{5_{3}}$&  $Q^{\prime}_{5_{4}}$ &  $Q^{\prime}_{5_{5}}$ &  $Q^{\prime}_{5_{6}}$ &  $Q^{\prime}_{5_{7}}$                                                                                               \\\hline\hline
\textbf{D1}, \textbf{D11} & 3/4 & -1/2 & 3/4 & -7/4 & -3/2 & -1/4 & -3/2 & 1 & -1/4 & 9/4 & 1 \\
\textbf{D2}, \textbf{D9} & 3/4 & -1/2 & -7/4 &3/4 & -3/2 & -1/4 & 1 & -3/2 & 9/4 & -1/4 & 1 \\
\textbf{D3}, \textbf{D6} &-3/4 & 7/4 & 1/2 & -3/4 & 3/2 & -1 & 1/4 & 3/2 & -9/4 & -1 & 1/4 \\
\textbf{D4}, \textbf{D10} &3/4 & -7/4 & 3/4 & -1/2 & -3/2 & 1 & -3/2 & -1/4 & 1 & 9/4 & -1/4 \\
\textbf{D5}, \textbf{D12} & -3/4 & -3/4 & 1/2 & 7/4 & 3/2 & 3/2 & 1/4 & -1 & 1/4 & -1 & -9/4 \\
\textbf{D7}, \textbf{D8} & -3/4 & -3/4 & 7/4 & 1/2 & 3/2 & 3/2 & -1 & 1/4 & -1 & 1/4 & -9/4 \\
 \hline\hline
	\end{tabular}
	\caption{ \small{$U(1)^{\prime}$ charges of Class D models. }}
	\label{tab:classD_charges}
	\end{table}

\begin{table}[H]
\centering\resizebox{\textwidth}{!}{%
\begin{tabular}{l|cccccccccccc|cccccc}\hline\hline
\textbf{Class D} & \multicolumn{12}{c|}{Multiplicities}                                   & \multicolumn{6}{c}{Charges$\times{\sqrt{10}}$}        \\\hline
Models  & $M_{13}$ & $M_{14}$ & $M_{15}$ & $M_{34}$ & $M_{35}$ & $M_{45}$ & $M_{31}$ & $M_{41}$ & $M_{51}$ & $M_{43}$ & $M_{53}$ & $M_{54}$ & $Q_{13}^{\prime}$ & $Q_{14}^{\prime}$ & $Q_{15}^{\prime}$ & $Q_{34}^{\prime}$ & $Q_{35}^{\prime}$ & $Q_{45}^{\prime}$  \\\hline\hline
\textbf{D1}, \textbf{D11}      &  1 & 1 & 3 & 4 & 1 & 2 & 2 & 2 & 1 & 1 & 4 & 1 &$\frac{5}{4}$ & 0 & $\frac{5}{2}$ & $-\frac{5}{4}$ & $\frac{5}{4}$ & $\frac{5}{2}$   \\
\textbf{D2}, \textbf{D9}  & 1 & 1 & 1 & 1 & 4 & 1 & 3 & 1 & 2 & 3 & 1 & 4  &$\frac{5}{4}$ & $-\frac{5}{2}$ & 0 & $\frac{5}{4}$ & $-\frac{5}{4}$ & $-\frac{5}{2}$   \\
\textbf{D3}, \textbf{D6} & 3 & 1 & 1 & 3 & 1 & 3 & 1 & 1 & 1 & 1 & 3 & 1 &$-\frac{5}{2}$ & $-\frac{5}{4}$ & 0 & $\frac{5}{4}$ & $\frac{5}{2}$ & $\frac{5}{4}$   \\
\textbf{D4}, \textbf{D10} & 3 & 1 & 1 & 1 & 3 & 1 & 1 & 1 & 1 & 3 & 1 & 3 &$\frac{5}{2}$ & 0 &  $\frac{5}{4}$ & $-\frac{5}{2}$ & $-\frac{5}{4}$ & $\frac{5}{4}$   \\
\textbf{D5}, \textbf{D12} & 1 & 1 & 2 & 1 & 3 & 1 & 1 & 4 & 1 & 3 & 1 & 3 &0 & $-\frac{5}{4}$ &  $-\frac{5}{2}$ & $-\frac{5}{4}$ & $-\frac{5}{2}$ & $-\frac{5}{4}$   \\
\textbf{D7}, \textbf{D8} & 1 & 2 & 1 & 2 & 1 & 4 & 3 & 1 & 3 & 1 & 3 & 1 & 0 & $-\frac{5}{2}$ &  $-\frac{5}{4}$ & $-\frac{5}{2}$ & $-\frac{5}{4}$ & $\frac{5}{4}$ \\\hline\hline
\end{tabular}}
\caption{ \small{Singlets spectrum of Class D models. }}
\label{tab:classD_singlets}
\end{table}

Phenomenological analysis of Model D9 was presented in the main body of the present text. 

Regarding the singlet sector of the models, their superpotential  can be written as

	\begin{equation}
	W\supset{\mu_{ij}^{\alpha\beta}}\theta^{\alpha}_{ij}\theta^{\beta}_{ji}+\lambda_{ijk}^{\alpha\beta\gamma}\theta_{ij}^{\alpha}\theta^{\beta}_{kj}\theta_{ki}^{\gamma}
	\end{equation}

\noindent where $\mu_{ij}^{\alpha\beta}$ are mass parameters and $\lambda_{ijk}^{\alpha\beta\gamma}$ dimensionless coupling constants. The greek indices run from 1 up to the multiplicitie $M_{ij}$ of the corresponding singlet. Minimalization of the superpotential ($\partial{W}/\partial{\theta_{ij}^{\alpha}}=0$) leads to the F-flatness conditions.

\section{Flavour violation bounds for the various models}\label{appC}

In the main text we have analyse in detail the low energy implications of model D9. A similar phenomenological analysis have been performed for all the MSSM spectrum models discussed so far. Due to the large number of models we do not present in detail the analysis for each model. Here we discuss the main flavor violation results for the four classes of MSSM models presented in the previous sections. 

Models of the same class share common $U(1)^{\prime}$ properties and consequently their phenomenological analysis is very similar. Next, we discuss the basic flavour violation bounds for each class of models. The main results collectively presented in Table \ref{tab:Zprimebounds}.

\textbf{Class A:} The six models that compromised the Class A  have very similar $U(1)^{\prime}$ charges. More specifically, only two values allowed for the $|Q^{\prime}|$ charges, $0$ and $1/2$. Matter fields descending from the $SU(5)$ tenplets have zero charge and as a result the corresponding flavor violation process appear very suppressed. The $Q^{\prime}$ charges appear (semi) non-universal in the lepton sector but again the corresponding LFV process are well suppressed in comparison with the experimental results. In summary, flavor violation process in Class A models appear to be suppressed and consequently  $M_{Z^{\prime}}$ bounds cannot extracted for this class of models.

\textbf{Class B:} From the twenty-four models of this class, eight-teen of them have been analysed in detail. In particular, the models B4, B5, B8, B13, B15 and B16  predict inappropriate mass hierarchies and as a result have been excluded from further analysis. For the remaining realistic models, the dominant constraints descents from the Kaon oscillation system. Approximately, the $Z^{\prime}$ contribution to the $K^{0}-\overline{K^{0}}$ mass split is

\begin{equation}
\Delta M_{K}^{Z^{\prime}}\simeq\frac{10^{-13}g^{\prime 2}}{M^{2}_{Z^{\prime}}} 
\end{equation}

\noindent which compared to the experimental bounds, for $g^{\prime}=0.5$ gives the constraint: $M_{Z^{\prime}}\gtrsim{190}$ TeV.

\textbf{Class C:} Due to the flux integers which characterize this class of models (see Table \ref{tab:classC_fluxes}), all the matter fields descending from the $SU(5)$ tenplets have the same $U(1)^{\prime}$ charges and as a result the corresponding flavour violation processes (like semi-leptonic meson decays  and meson mixing effects) are suppressed. However, on the lepton sector the $U(1)^{\prime}$ charges are non-universal leading to lepton flavor violation phenomena at low energies. The dominant constraint descent from the three body decay $\mu^{-}\rightarrow{e^{-}e^{-}e^{+}}$. Approximately for all the C-models, we find that the $Z^{\prime}$ contributions to the branching ratio of the decay is

\begin{equation*}
Br(\mu^{-}\rightarrow{e^{-}e^{-}e^{+}})\simeq{7.2\times{10^{-6}}\left(\frac{g^{\prime}\;\;\mathrm{TeV}}{M_{Z^{\prime}}}\right)^{4}}
\end{equation*}  

\noindent which compared to the current experimental bound implies that $M_{Z^{\prime}}\gtrsim{(51.8\times{g^{\prime}})}$  TeV, where $g^{\prime}$ the $U(1)^{\prime}$ gauge coupling. In the absence of any signal in future $\mu^{-}\rightarrow{e^{-}e^{-}e^{+}}$ searches, this bound is expected to increased by one order of magnitude: $M_{Z^{\prime}}\gtrsim{(518\times{g^{\prime}})}$ TeV.

\textbf{Class D:} In this class of models the dominant constraints descend from the Kaon system. In some cases, strong bounds will be placed by future $\mu^{-}\rightarrow{e^{-}e^{-}e^{+}}$ searches. In particular,  for the models D1, D2, D5, D6, D8 and D10 the constraints from $Z^{\prime}$ contributions to the $K^{0}-\overline{K^{0}}$ mass split is: $M_{Z^{\prime}}\gtrsim{(475\times{g^{\prime}})}$ TeV. For the rest of D-models (D3, D4, D7, D9, D11, D12), the results are similar with those of model D9 which have been detailed analysed in the main body of the present text.

\begin{table}[H]
\centering\resizebox{\textwidth}{!}{%
\begin{tabular}{l|c|c}\hline\hline
\textbf{Models} & \textbf{Dominant Process} & $(M_{Z^{\prime}}/g^{\prime})$ \textbf{bound} (TeV)
 \\\hline
 Class-B &   $K^{0}-\overline{K^{0}}$ mixing & $M_{Z^{\prime}}/g^{\prime}\gtrsim{380}$\\
(excluded: B4, B5, B8, B13, B15, B16) & &\\\hline
 & $\mu^{-}\rightarrow{e^{-}e^{-}e^{+}}$  &  $M_{Z^{\prime}}/g^{\prime}\gtrsim{51.8}$ \\
 Class-C & & \\
 & Future $\mu^{-}\rightarrow{e^{-}e^{-}e^{+}}$ searches & $M_{Z^{\prime}}/g^{\prime}\gtrsim{518}$ \\\hline
 D1, D2, D5, D6, D8, D10 & $K^{0}-\overline{K^{0}}$ mixing &  $M_{Z^{\prime}}/g^{\prime}\gtrsim{475}$\\\hline
  & $K^{0}-\overline{K^{0}}$ mixing &  $M_{Z^{\prime}}/g^{\prime}\gtrsim{238}$\\
  D3, D4, D7, D9, D11, D12 & & \\ 
  & Future $\mu^{-}\rightarrow{e^{-}e^{-}e^{+}}$ searches & $M_{Z^{\prime}}/g^{\prime}\gtrsim{420}$\\\hline\hline
\end{tabular}}
\caption{ \small{Dominant flavour violation process for each model along with the corresponding bounds on the mass of the  flavour mixing $Z^{\prime}$ boson. }}
\label{tab:Zprimebounds}
\end{table}

\end{appendices}

\end{document}